\documentclass[10pt, final, journal, letterpaper, oneside, twocolumn]{IEEEtran}
\usepackage{cite}
\usepackage{amsmath,amssymb,amsfonts}
\usepackage{algorithmic}
\usepackage{graphicx}
\usepackage{textcomp}
\usepackage{float}
\usepackage{subfloat}
\usepackage{caption}
\usepackage{subcaption}

\DeclareMathAlphabet\mathbfcal{OMS}{cmsy}{b}{n}

\begin{document}
\title{Statistical Characterization of Wireless MIMO Channels in Mode-Stirred Enclosures}
\author{Mir~Lodro, \IEEEmembership{Member,~IEEE}, Steve~Greedy, Sirio~Belga~Fedeli, Christopher~Smartt, Ana~Vukovic, David~Thomas and Gabriele~Gradoni, \IEEEmembership{Member,~IEEE} \vspace{-0.75cm}
\thanks{Mir Lodro, Steve Greedy, Christopher Smartt, Ana Vukovic, David Thomas and Gabriele Gradoni are with George Green Institute for Electromagnetic Research-GGIEMR, the University of Nottingham, UK. Gabriele Gradoni is also with British Telecommunication and the University of Cambridge, UK. Sirio Belga Fedeli is with Department of Mathematics, King's College London, UK}}
\maketitle
\begin{abstract}
We present the statistical characterization of a 2x2 Multiple-Input Multiple-Output wireless link operated in a mode-stirred enclosure, with channel state information available only at the receiver (agnostic transmitter). Our wireless channel measurements are conducted in absence of line of sight and varying the inter-element spacing between the two antenna elements in both the transmit and receive array. The mode-stirred cavity is operated: i) at a low number of stirrer positions to create statistical inhomogeneity; ii) at two different loading conditions, empty and with absorbers, in order to mimic a wide range of realistic equipment level enclosures. Our results show that two parallel channels are obtained within the confined space at both the operating conditions. The statistical characterization of the wireless channel is presented in terms of coherence bandwidth, path loss, delay spread and Rician factor, and wideband channel capacity. It is found that the severe multipath fading supported by a highly reflecting environment creates unbalance between the two Multiple-Input Multiple-Output channels, even in presence of substantial losses. Furthermore, the channel capacity has a multi-modal distribution whose average and variance scale monotonically with the number of absorbers. 
Results are of interest in IoT devices, including wireless chip-to-chip and device-to-device communications, operating in highly reflective environments. 
\end{abstract}
\vspace{-0.17cm}
\begin{IEEEkeywords}
SDR, USRP, MIMO, IoT, Mode-Stirred Enclosure, Channel Characterization.
\end{IEEEkeywords}
%
%
%
%
\vspace{-0.35cm}
\section{Introduction}\label{sec:introduction}
Statistical channel characterization and modeling is one of the foremost step when Multiple-Input Multiple-Output (MIMO) wireless systems are to be deployed in complex and highly reflective radio propagation environments. In scanarios where small devices communicate with compact anetnna arrays in confined environments, a situation occurring for example in chip-to-chip and device-to-device communications, the role of multipath fading becomes increasing important and traditional channel models fail to descirbe the fading statistics. Measurement campaigns for statistical characterization and modeling of compact MIMO systems operating within confined environments deserve more attention, especially in view of design and roll out of Internet-of-Things networks. The MIMO wireless channel characterization can be tackled from different perspectives that depends on various factors such as type and number of transmit and receive antennas, their heights, polarization, and the layout of the propagation channel. The wireless channel is normally influenced by the number of scatterers, their material profile, as well as their distribution around transmitter and receiver. Remarkably, in \cite{chen2007inter} different approaches have been used to study a Single-Input Single-Output (SISO) chip-to-chip (C2C) channels in computer casing. Other Investigators \cite{zajic2018modeling}\cite{fu2019300}\cite{kim2016300} have presented two ring models for stationary wireless C2C channel in metal enclosure with a SISO system. 
The study in \cite{karedal2007characterization}\cite{redfield2011understanding} show measurements for ultra-wideband (UWB) board-to-board communication within computer chassis, and found a uniform propagation environment independent of transmit/receive antenna location, with small variations in channel path gains. Further, in \cite{khademi2015channel} the Investigators have performed channel measurements in metal cabinet and in \cite{ohira2011experimental}\cite{nakamoto2013wireless} they have performed channel measurements in ICT equipment. 

In this work, a metal mode-stirred enclosure is used to emulate a complex MIMO propagation channel. Additionally, the metal enclosure is modified by rotating an irregular stirrer inside the metal enclosure, under different loading conditions. This provides a means to emulate a dynamical multi-path channel supported by an highly-reflective environment. This study complements the recent study of low-dimensional MIMO systems operated in a large scale dynamical environment emulated by a reverberation chamber operated with a live base station \cite{Micheli2021}.
Our observations point towards the need of abetter understanding of the fundamental propagation mechanisms in complex electromagnetic environments, as well as their relation with information theoretic metrics to evaluate the achievable performance of future wireless communication systems. 

After the introduction in Section I, Section II. explains parameter baseband model for MIMO channel measurements and parameter extraction. SDR based MIMO measurement setup in explained in Section III. In Section IV., we presented key statistical parameters of MIMO channel such as delay spread, coherence bandwidth, pathloss, Rician factor, and channel capacity statistics. This completes the statistical characterization of these parameters. In Section V. we draw some Conclusion and Future Perspective.

\vspace{-0.35cm}
\section{SDR Based Parameter Extraction for Virtual MIMO}
The MATLAB/Simulink framework is used as a baseband waveform development platform to extract the channel transfer function parameters. We use a virtual MIMO setting to measure MIMO channel transfer matrix through a software-defined-radio (SDR) platform. The virtual MIMO array measurements are suitable for wireless environments with fixed antenna arrays. The scattering parameters are measured using a frequency-sweep approach where the local oscillators of the transmit and receive Universal Software Radio Peripherals (USRPs), both X310, are synchronously tuned from 5.50 GHz to 5.70 GHz. The center frequency of the local oscillators of the transmit and receive USRPs was also tuned synchronously. The sweep signal generated in Simulink is updated every 0.025 seconds and the local oscillator stability time was 0.1 seconds. Therefore, at every frequency there were four snapshots of $N=10,000$ long complex samples and the total time for one frequency sweep was $0.1*N=50.1$ seconds excluding from Simulink initialization and start-up time. 
Baseband sampling rate was set to 400 Ksps and the RF front-end sampling rate was also 400 ksps. The two X310 USRPs received 10 MHz and 1 PPS reference signals from Octoclock-G using equal length cables. We recorded both complex I and Q samples per SDR channel as well as received power per SDR channel along with the metadata information of each measurement.

\begin{figure*}
    \centering
    \includegraphics[width=\textwidth]{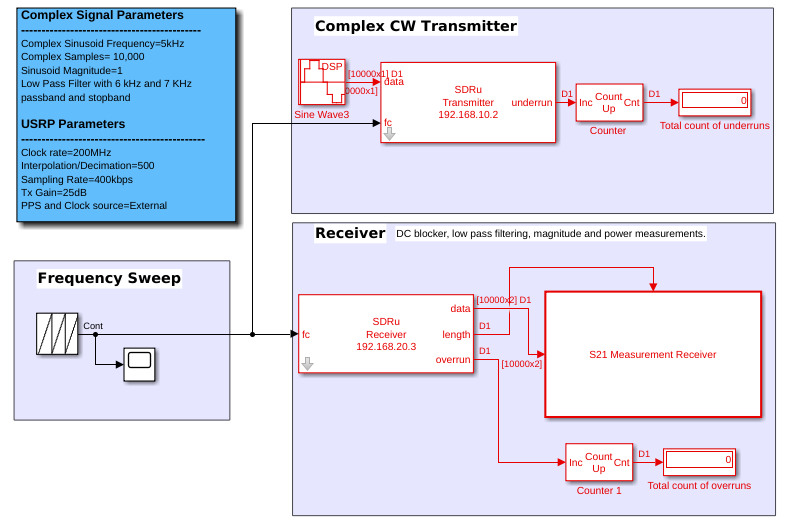}
    \caption{Simulink model for the MIMO frequency transfer function measurement.}
    \label{fig:baseband_2x2_virtual}
\end{figure*}

\begin{figure*}
    \centering
    \includegraphics[width=\textwidth]{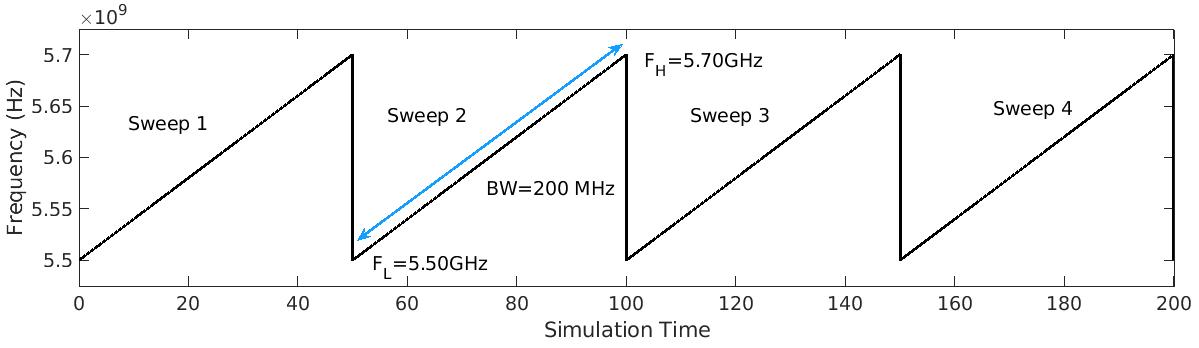}
    \caption{Number of frequency sweeps.}
    \label{fig:sweeps}
\end{figure*}

\begin{figure}
        \centering
        \includegraphics[width=0.45\textwidth]{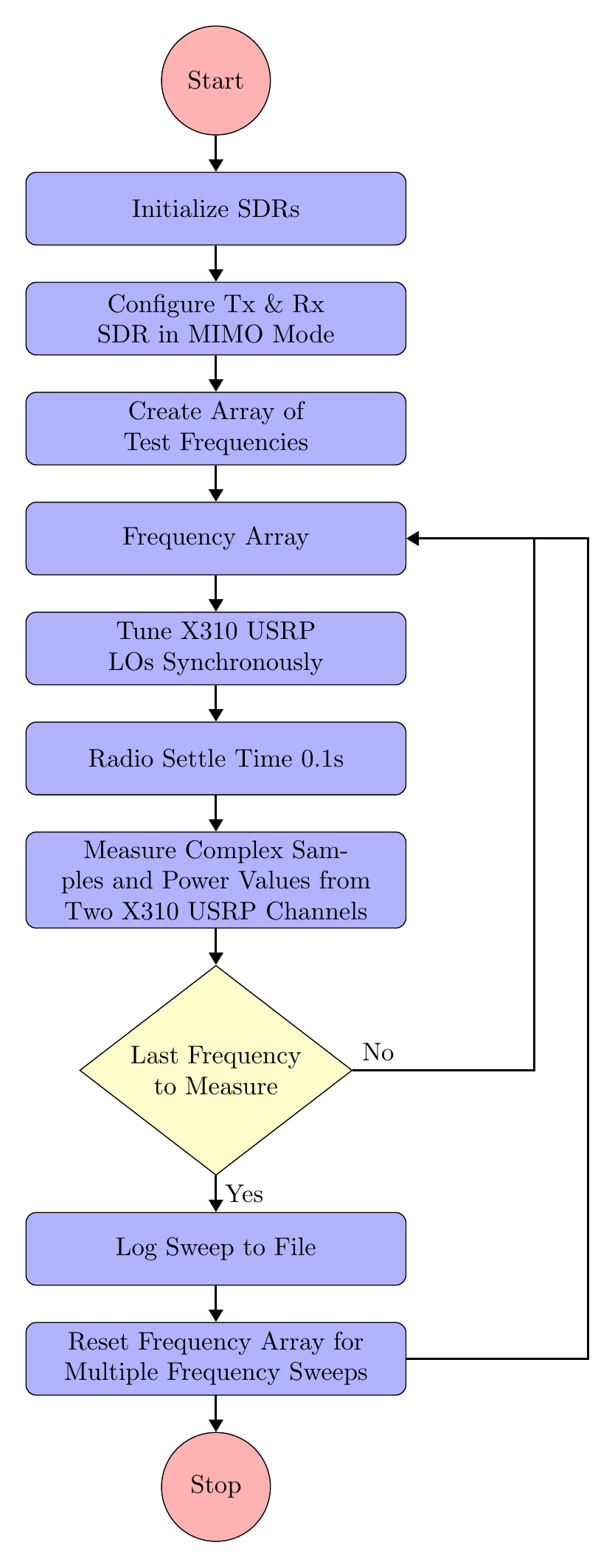}
        \caption{USRP based channel measurement flowchart.}
        \label{fig:tf_flowchart}
    \end{figure}

The frequency transfer function is measured by sweeping the local oscillators of the two USRP synchronously. Figure \ref{fig:baseband_2x2_virtual} shows the baseband model to perform the frequency sweep followed by the measurement of the received signal. It was assured that there is no underflow and overflow during the measurement cycle, which is important in MIMO channel characterization. This model is thus used to perform MIMO measurements using virtual MIMO technique where the channel is measured in two phases. In the first phase the instant channel between transmit antenna 1 and receive antenna 1 and receive antenna 2 is measured, in the second phase the instant channel between transmit antenna 2 and receive antenna 1 and receive antenna 2 is measured. 
Figure \ref{fig:sweeps} shows four frequency sweeps, each of 200 MHz bandwidth. The full MIMO channel measurement can be obtained with the same baseband model without switching transmitting channels. This can be acheived by transmitting two baseband complex Sinusoids at different frequencies while sweeping local oscillators synchronously. In the baseband model, both the lowpass filter and the bandpass filter can be employed to measure the magnitude of the two received Sinusoids.
The scattering parameter measurement steps are presented in the flowchart \ref{fig:tf_flowchart}.
\vspace{-0.35cm}
\section{Measurement Setup}
\begin{figure}
    \centering
    \includegraphics[width=\linewidth]{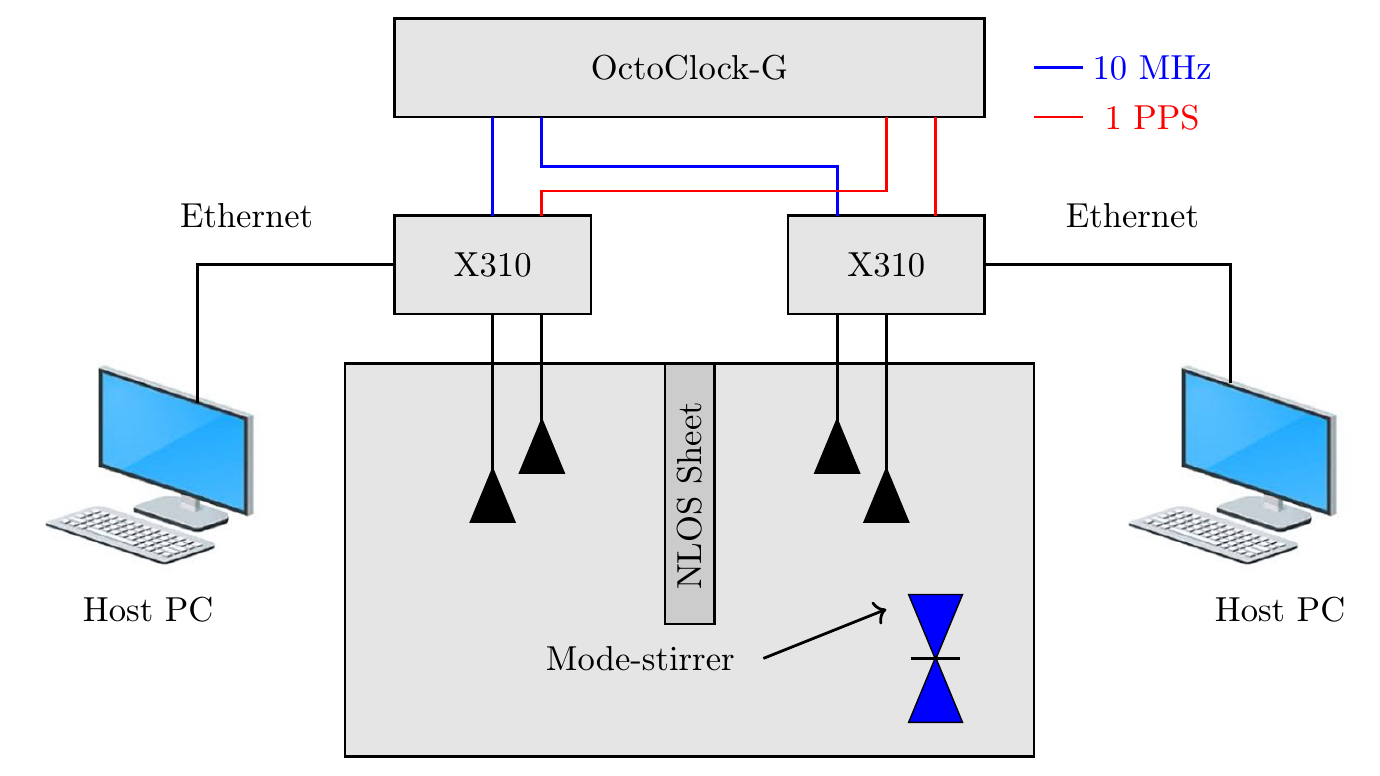}
    \caption{Complete sketch of the no-line-of-sight MIMO channel measurement setup inside a metal enclosure.}
    \label{fig:mimo_sounder_sketch_nlos}
\end{figure}

\begin{figure}
    \centering
    \includegraphics[width=0.45\textwidth]{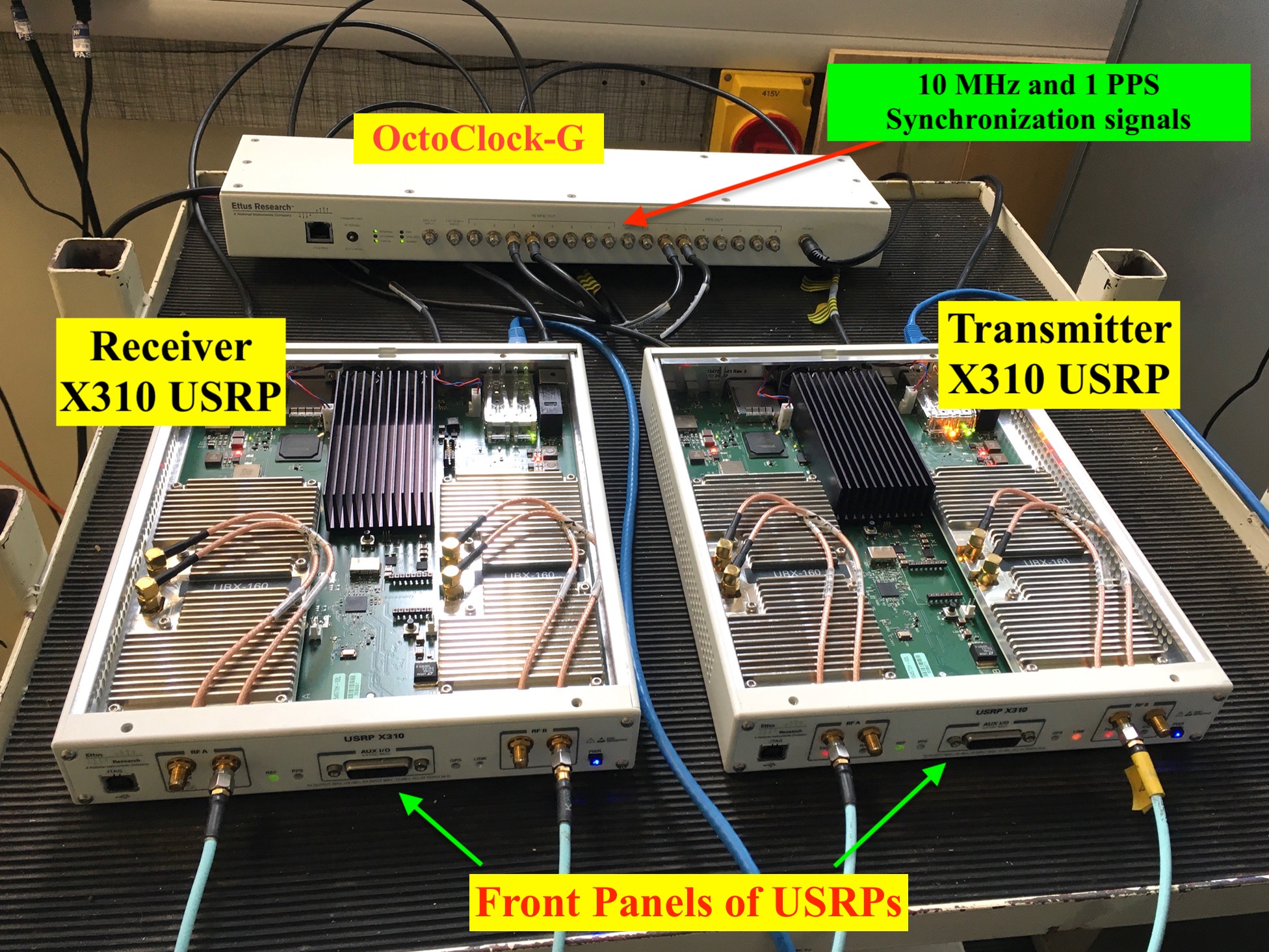}
    \caption{Front view of the 2x2 MIMO measurement setup based on two X310 USRPs.}
    \label{fig:mimo_x310_front}
\end{figure}

Compact MIMO wireless systems, including C2C communication paltforms \cite{rayess2017antennas}, can be tested and characterized in the GHz regime by employing our measurement setup combining SDR devices with a small reverberation chamber. 
Here, the SDR based test-bed was used to obtain complex frequency transfer function of the MIMO wireless C2C channel. The frequency was swept from 5.50GHz to 5.70GHz. 
The channel measurements were performed with four identical and vertically polarized monopole antennas connected with input and output ports of the SDR using RF cables.
Figure \ref{fig:mimo_sounder_sketch_nlos} shows the sketch of the USRP-based 2x2 MIMO measurement setup. The two USRPs are connected to independent host Personal Computers for signal generation and transmission over the fading channel emulated in the metal enclosure. This is important aspect to avoid inter-board or intra-board RF interference. A significant channel leakage was observed when the 2x2 MIMO channel was measured using same X310 USRP. Therefore, this type of setup gives general assessment of compact MIMO channels but antenna cabling needs to be handled carefully. Inherently, a 2x2 MIMO system using independent USRPs as transmitter and receiver could also be developed by connecting both USRPs to single high-performance PC. This will require host PC to have dual NIC installed or one 10G Ethernet card with two interface ports.  In this way, data acquisition is performed on the same PC and packet mismatched can be avoided because of different baseband model runtimes, and hence increasing error vector magnitude (EVM) and bit error rate (BER) measurement accuracy for wireless communication system measurements. Again, in our solution the channel transfer function is measured by sweeping the local oscillator frequency of transmit and receive USRP simultaneously over a specified bandwidth.
Figure \ref{fig:mimo_x310_front} shows the front view image of an USRP based 2x2 MIMO testbed. It shows the two X310 USRPs front panels and the two UBX-160 RF daughterboards per USRPs. 
The Octoclock-G is used to supply 10 MHz and 1 PPS synchronization signals to both the USRPs. In order to avoid leakage and the effect of intra-board and inter-board interference, separate UBX-160 RF cards are used for each transmit and receive channels.

\begin{figure}
\centering
\subfloat[]{\label{fig:mimo_cloud}\includegraphics[width=0.25\textwidth]{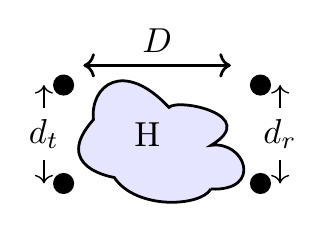}}
\subfloat[]{\label{fig:mimo_general}\includegraphics[width=0.25\textwidth]{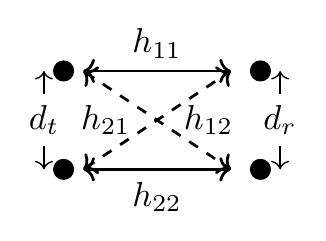}}
\caption{MIMO Configuration (a) MIMO channel matrix $\mathbf{H}$ (b) MIMO channel elements.}
\label{fig:mimo_c2c_sketch}
\end{figure}

Figure \ref{fig:mimo_c2c_sketch} shows the sketch of the 2x2 MIMO channel measurement setup, where transmitter and receiver units are located at distance $D$ apart. Furthermore, the transmitter and receiver elements are separated by distances $d_t$ and $d_r$ respectively.

\begin{figure}
    \centering
\subfloat[]{\includegraphics[width=0.49\linewidth]{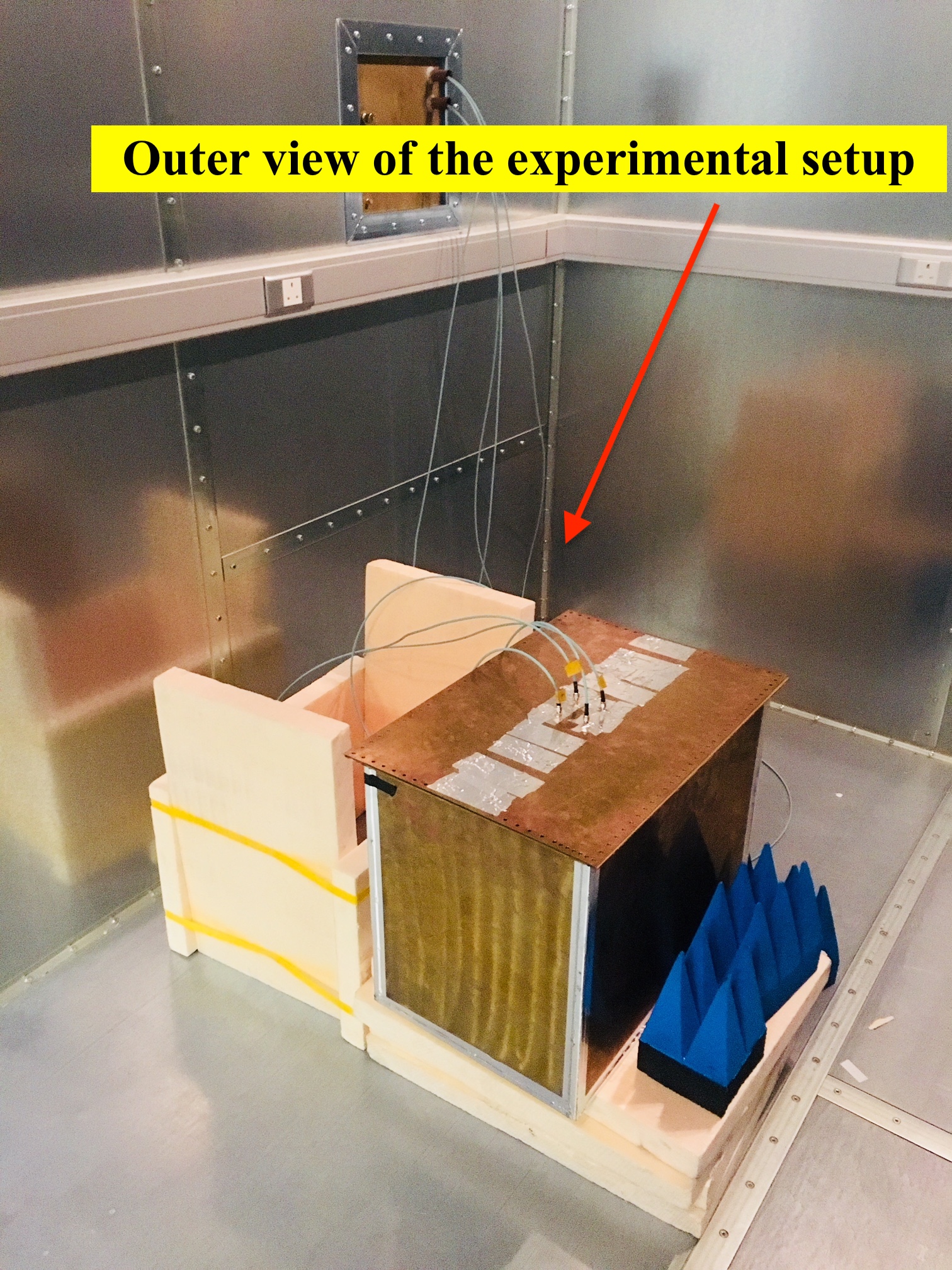}}
\hfill
\subfloat[]{\includegraphics[width=0.49\linewidth]{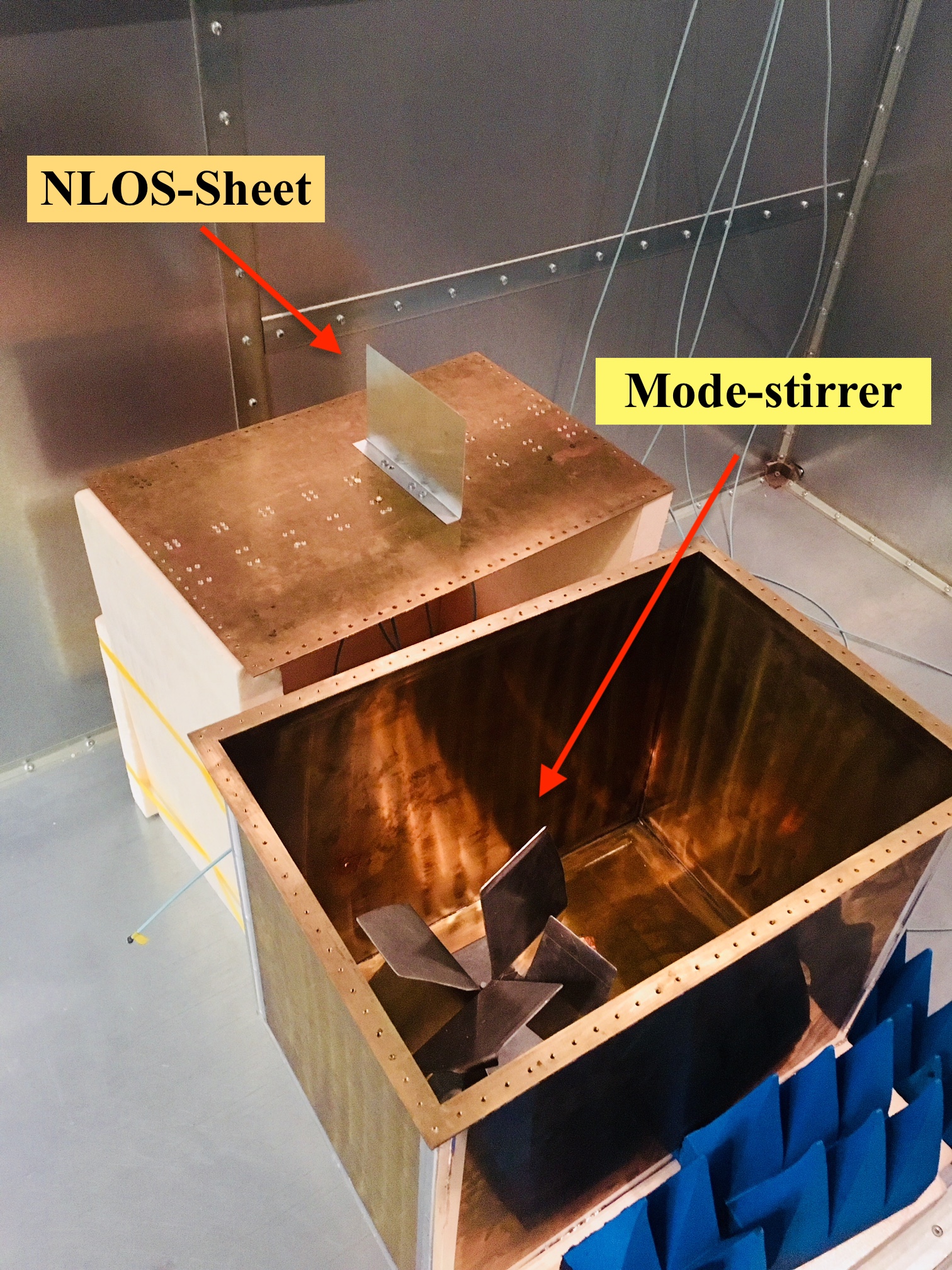}}\\
\subfloat[]{\includegraphics[width=0.49\linewidth]{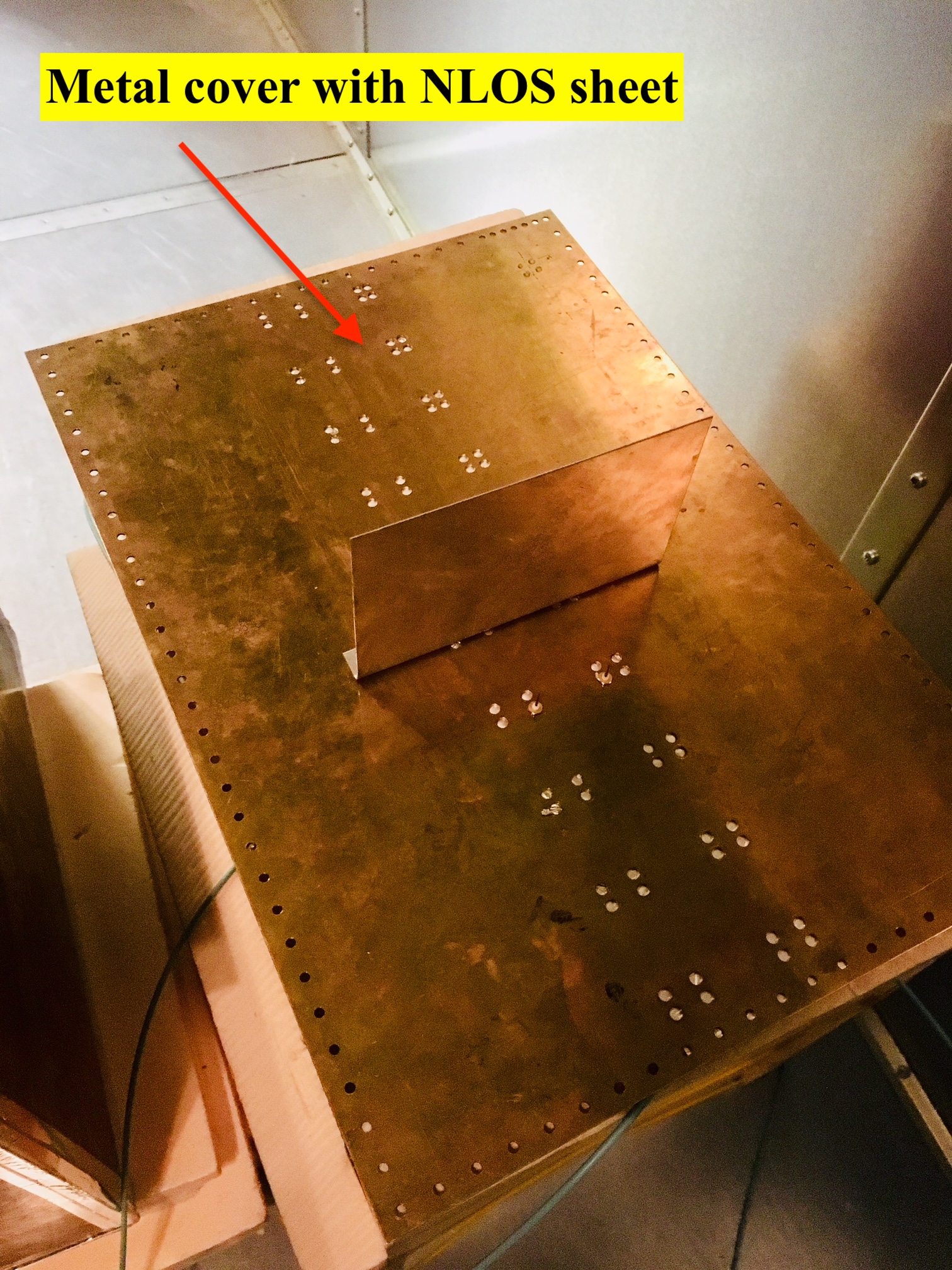}}
\hfill
\subfloat[]{\label{fig:mode_stirrer}\includegraphics[width=0.49\linewidth]{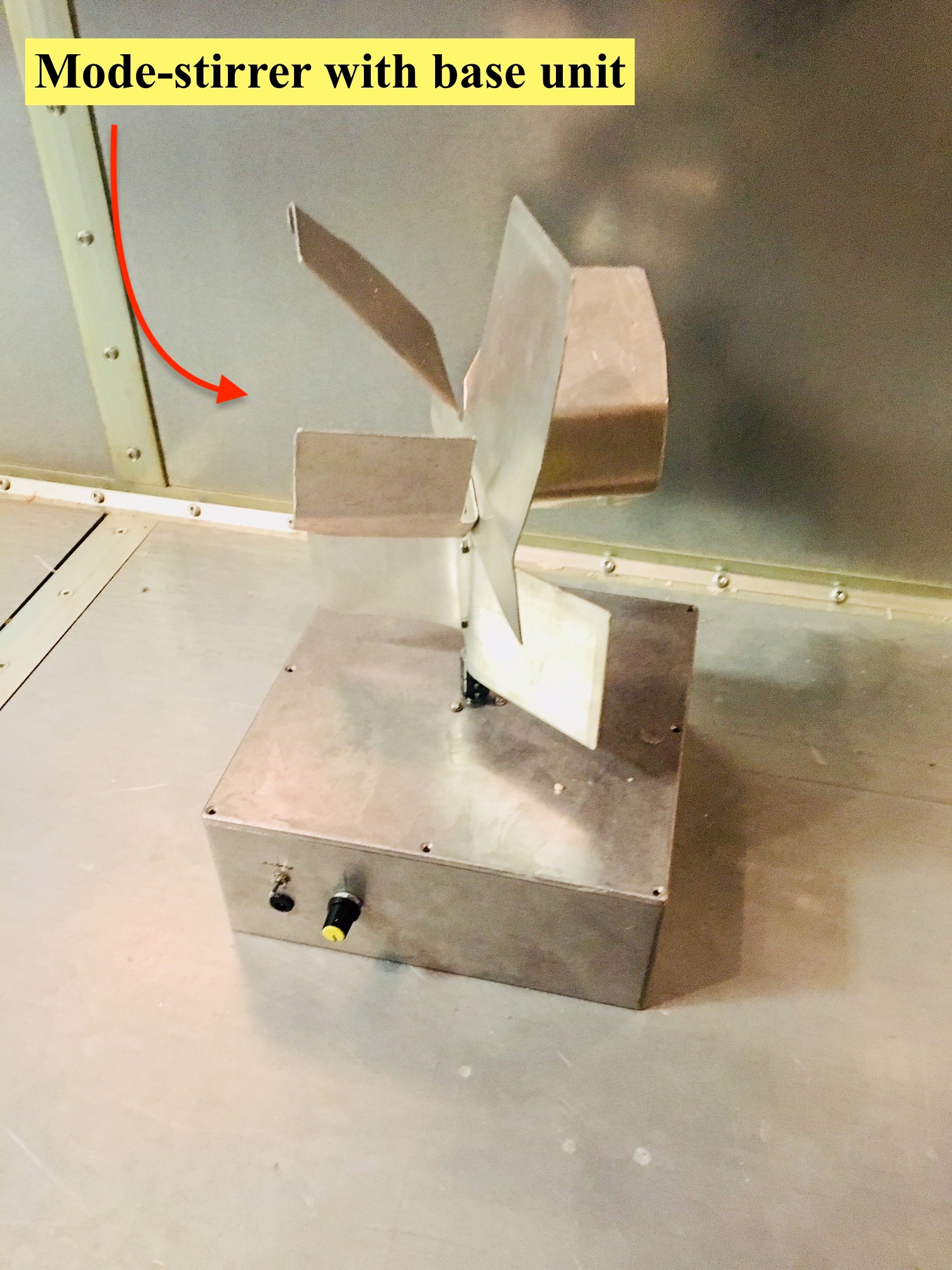}}
\caption{Measurement setup: (a) Overview of measurement setup (b) Inside view of measurement setup showing panel and inside view of the metal enclosure (c) Metallic panel with a conducting sheet creating the no-line-of-sight (d) Asymmetric mode-stirrer.}
\label{fig:meas_setup}
\end{figure}

We used the USRP X310 platforms from Ettus research for our wideband channel sounder to perform wideband measurements using four similar vertically polarized monopole antennas. 
Inside the enclosure, we also used a metal sheet of dimension $13\,cm \times 20\,cm$ to create the MIMO no-line-of-sight (NLOS) propagation environment. This form of channel arrangement is known to have increased MIMO channel capacity as it reduces the possibility of generating rank-deficient or degenerate MIMO channels. 
In other words, it creates a rich multi-path fading environment which is supposed to maximize the MIMO channel capacity. The metal enclosure has dimensions of $h\times \ell \times w$ of $45\,cm\times 37\,cm\times 55\,cm$ with a mechanical mode-stirrer in it as shown in Fig. \ref{fig:mode_stirrer}. The mode-stirrer was controlled using an external Yuasa battery. We performed the NLOS MIMO channel measurements in an empty metal enclosure first, and then in presence of losses when the metal enclosure was loaded with 2, 4 and 8 RF absorbers cones (normally employed to coat the interior of anechoic chambers). We conducted NLOS MIMO channel measurements for two sets of transmit and receive inter-element distances at ten different stirrer positions. Since the SDR sweep time was higher than the coherence time of the channel, we performed frequency response measurements at stationary stirrer positions. The measurements were repeated when the metal enclosure was empty and when it was sequentially loaded with two, four and eight RF absorbers cones. 
The value of the measurement parameters are summarized in Table .\ref{tab:measurement_setup}.
\begin{table}[h]
    \centering
     \caption{MEASUREMENT SYSTEM PARAMETERS}
    \begin{tabular}{|c|c|}
        \hline
        Parameters & Value  \\
        \hline
         Bandwidth & 200 MHz\\
         Measurement Points& 501\\
         Start Frequency & 5.50 GHz\\
         Stop frequency & 5.70 GHz\\
         Tx/Rx Polarization & Vertical\\
         Time Resolution & 5 ns\\
         Frequency Resolution & 400 KHz\\
         USRP clock rate & 200 Mbps\\
         Sample Rate & 400 Ksps\\
         USRP Decimation/Interpolation&500\\
         \hline
    \end{tabular}
   
    \label{tab:measurement_setup}
\end{table}
\vspace{-0.35cm}
\section{Measurement Results}

\subsection{MIMO System Model}
The discrete-time input output relationship of MIMO system with $N_T$ transmit antennas and $N_R$ receive antennas is given as:
\begin{equation}
    \mathbf{y=Hx+n},
\end{equation}
where $\mathbf{y}\in \mathbb{C}^{N_R\times 1}$ is the complex-valued received signal, $\mathbf{x}\in \mathbb{C}^{N_T\times 1}$ is the complex-valued transmitted signal, $\mathbf{n}\in \mathbb{C}^{N_R\times 1}$ is an additive Gaussian noise term, and $\mathbf{H}\in \mathbb{C}^{N_T\times N_R}$ is the MIMO channel matrix of dimensions $N_T \times N_R$ with complex-valued subchannel elements denoting path gains between $N_T$ transmit and $N_R$ receive antennas.

\subsection{Channel Capacity}
The maximum rate of information that a channel can support with arbitrary low probability of error is known as the channel capacity \cite{chuah2002capacity}. 
The wideband MIMO channel capacity over a bandwidth $B$ is given by \cite{molisch2002capacity}\cite{tse2005fundamentals}:
\begin{equation}
    C=\frac{1}{B}\int_{f_{min}}^{f_{max}}\log_2\left(1+\frac{\rho}{N_T} H(f)^HH(f)\right)\,df \quad bits/s/Hz.
\end{equation}
Equivalently, the channel capacity for measurements performed at discrete frequency points distributed across a range $[-B/2,B/2]$, is defined as:
\begin{equation}
    C=\frac{1}{B}\sum_{k=1}^{B}\log_2\left(1+\frac{\rho}{N_T} |H(f_k)|^2 \right)\quad bits/s/Hz,
\end{equation}
where $B$ is the frequency bandwidth, $H(f)$ is complex-valued wideband frequency transfer function of the MIMO channel and $\rho$ is the signal to noise ratio (SNR) per receive antenna. A $200 MHz$ channel bandwidth has been evaluated with 501 subchannels of $400 KHz$ each. For a proper MIMO system operation, the MIMO channel matrix should demonstrate useful properties. Firstly, it should not be rank deficient i.e. it shouldn't be subject to a reduction of the rank due keyhole or pin-hole channels. More precisely, the keyhole channels are known as degenerate MIMO channels whose rank is 1. In this condition, the capacity of degenerate MIMO channels is equal to that in SISO mode. Secondly, the MIMO channel matrix should exhibit a condition number $\geq 1$ or $0\,dB$. We remind that the rank of channel matrix $\mathbf{H}$ is the number of non-zero singular values. It indicates number of decodable data streams that can be spatially multiplexed across the wireless link. Operatively, the condition number of a MIMO channel matrix is defined as ratio of the maximum singular value $\sigma_{max}$ to minimum singular value $\sigma_{min}$:
\begin{equation}
    K(\mathbf{H})=\frac{\sigma_{max}}{\sigma_{min}}\geq 1, 
\end{equation}
which can be expressed in dB as:
\begin{equation}
    K(\mathbf{H})_{dB}=20\log_{10}\left( K(\mathbf{H})\right).
\end{equation}
We also remind that the singular values $\sigma_i$ of the channel matrix are related to the eigenvalues as $\lambda_i=\sigma_{i}^{2}$. Therefore, the condition number in terms of eigenvalues of MIMO channel matrix $\mathbf{H}$ is given as:
\begin{equation}
    K(\mathbf{H})_{dB}=10\log_{10}\frac{\lambda_{max}}{\lambda_{min}}\geq 0 dB.
\end{equation}

\begin{figure}
    \centering
    \includegraphics[width=\linewidth]{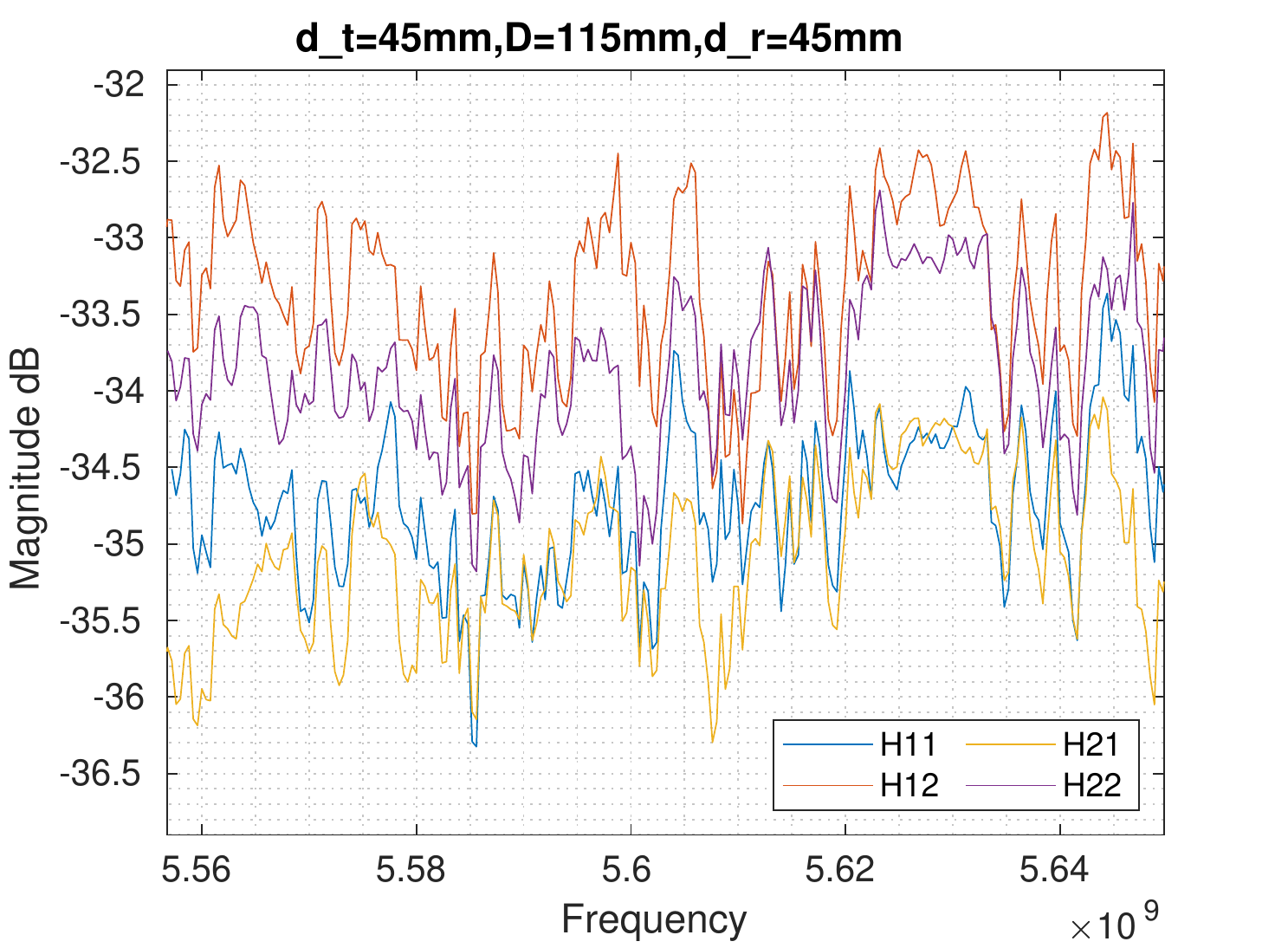}
    \caption{Complex transfer function of MIMO subchannels.}
    \label{fig:tf_subchannels}
\end{figure}

Figure \ref{fig:tf_subchannels} shows the transfer function of the MIMO sub-channel elements for our 2x2 wireless link. The transfer function shows that diversity is introduced by the highly reflective environment created by the metal enclosure. Additionally, it can be observed that the transfer function of subchannel elements has similar shape and different path loss profile in each channel element. 
This is expected as no buffer decoupling and matching network has been used to achieve equally strong channels \cite{Phang2018}. 

\begin{figure}
    \centering
    \includegraphics[width=\linewidth]{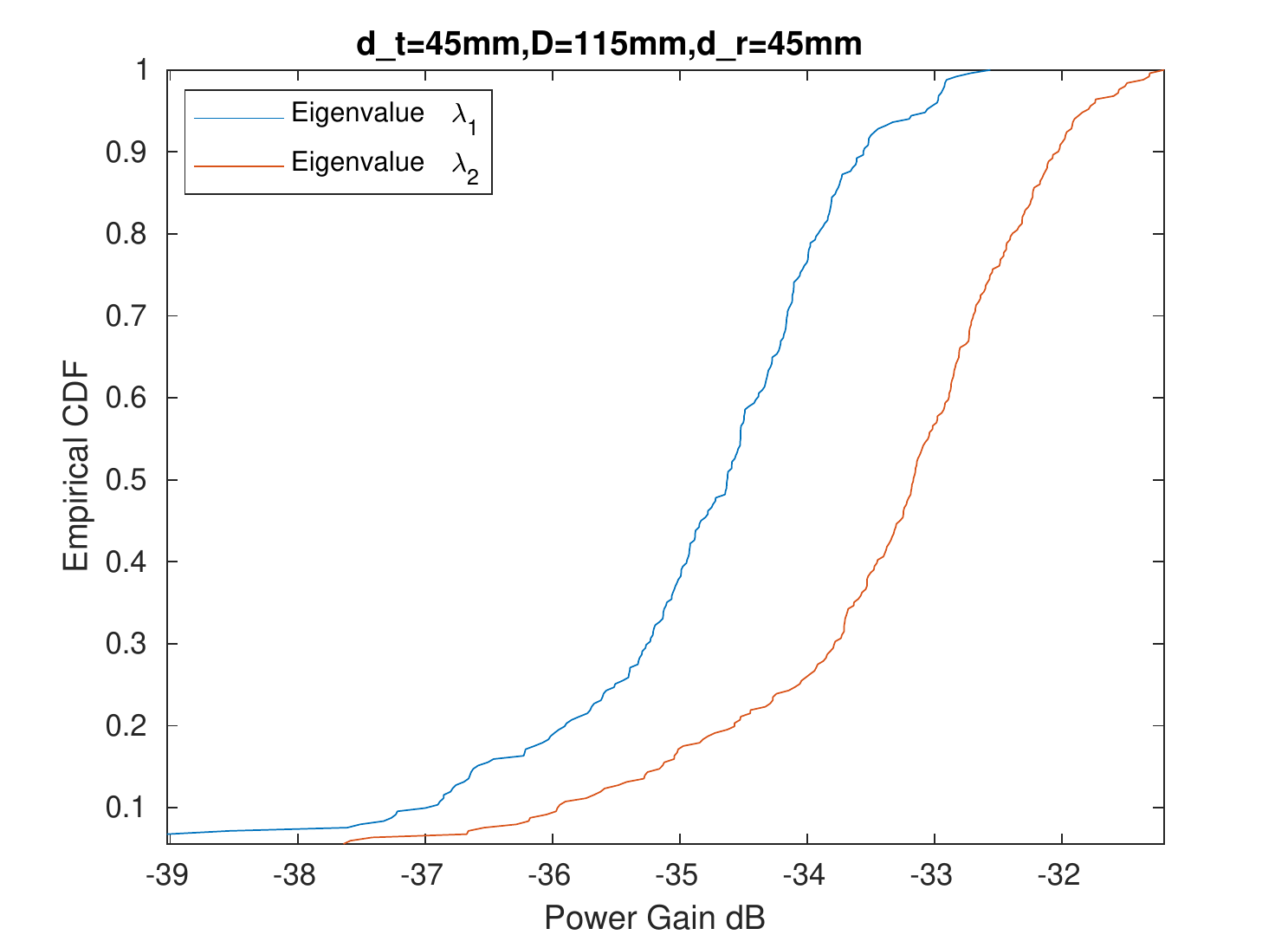}
    \caption{Eigenvalues of wireless chip-to-chip channel.}
    \label{fig:eigenvalues}
\end{figure}

Figure \ref{fig:eigenvalues} shows the empirical cumulative density function (CDF) of the eigenvalues of the MIMO wireless channel. It can be seen that the gain difference at $10\%$ probability is $1\,dB$. Therefore, at $10\%$ probability condition number is $1\,dB$ that meets the expected definition of a proper MIMO channel matrix.

\subsection{Coherence Bandwidth}
The coherence bandwidth is measured as the auto-correlation of the frequency transfer function $H(f)$. It is the bandwidth over which the signal fluctuations are considered flat. The coherence bandwidth is determined by measuring the width of frequency auto-correlation function at particular threshold usually above 0.5. It is generally helpful for the selection of the design parameters of wireless communications systems in any complex radio propagation environment that is unknown. This parameter can greatly affect the complexity of wireless C2C communication system as the inclusion and selection of equalizer as a design component for combating the inter-symbol interference (ISI) depends on this parameter \cite{TannerNEMF}. It is inversely related to RMS delay spread and coherence bandwidth, and inn terms of RMS delay spread at threshold of 0.5 is given by:
\begin{equation}
    B_{c}\approx\frac{1}{5\tau_{rms}},
\end{equation}
while at threshold of 0.9 reads: 
\begin{equation}
    B_{c}\approx\frac{1}{50\tau_{rms}}.
\end{equation}

\begin{figure}
    \centering
    \includegraphics[width=\linewidth]{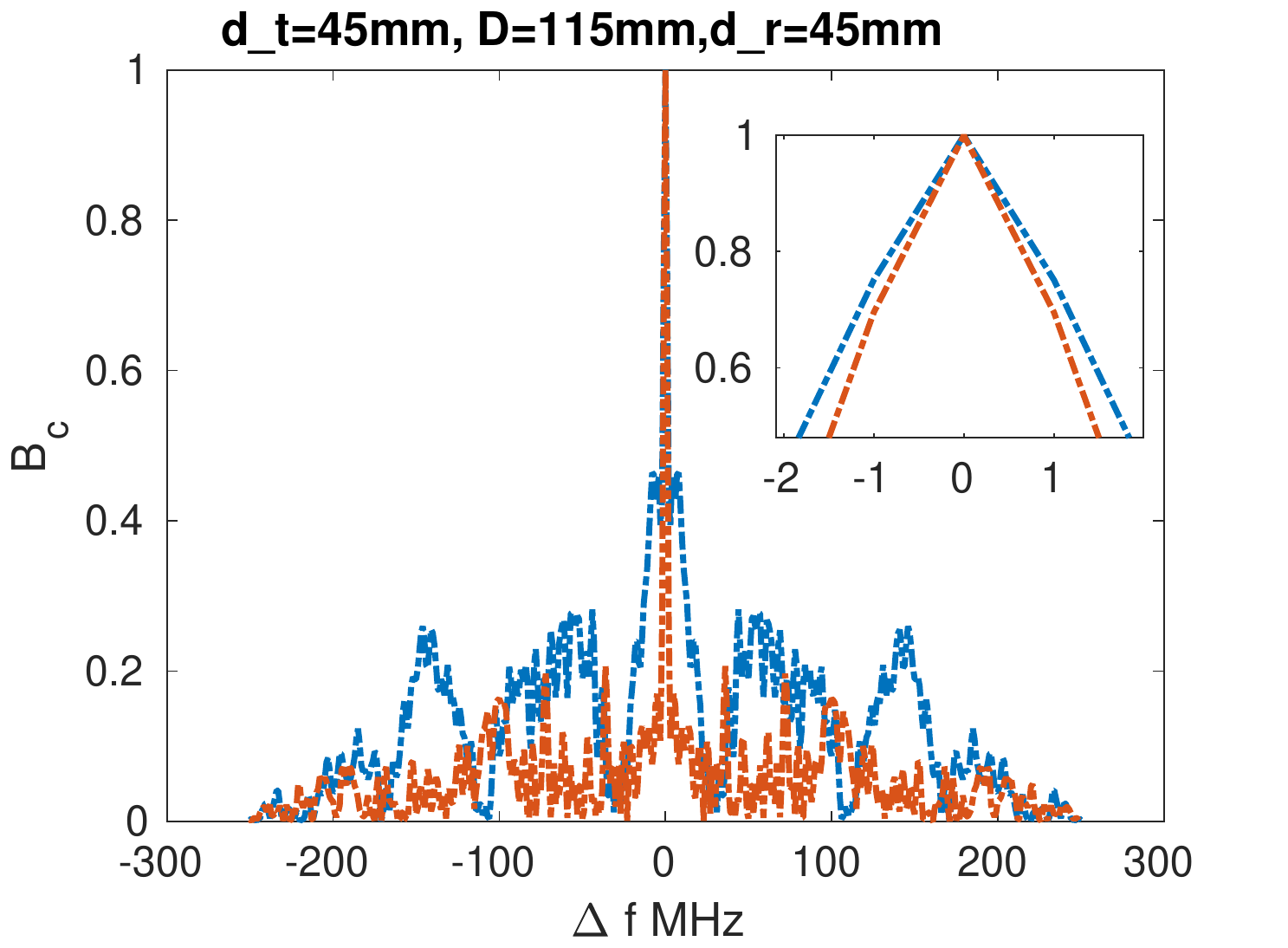}
    \caption{Coherence bandwidth of two SDR channels.}
    \label{fig:ch_bw}
\end{figure}

The coherence bandwidth can also be defined by using the concept of average mode-bandwidth in a loaded RC \cite{chen2011estimation}\cite{delangre2008delay}. 
Figure \ref{fig:ch_bw} shows the measured coherence bandwidth of the two MIMO channels. It is shown that the maximum coherence bandwidth value is $2\,MHz$.

\subsection{Path Loss Model}
\begin{figure}
    \centering
   \subfloat[]{\includegraphics[width=0.49\linewidth]{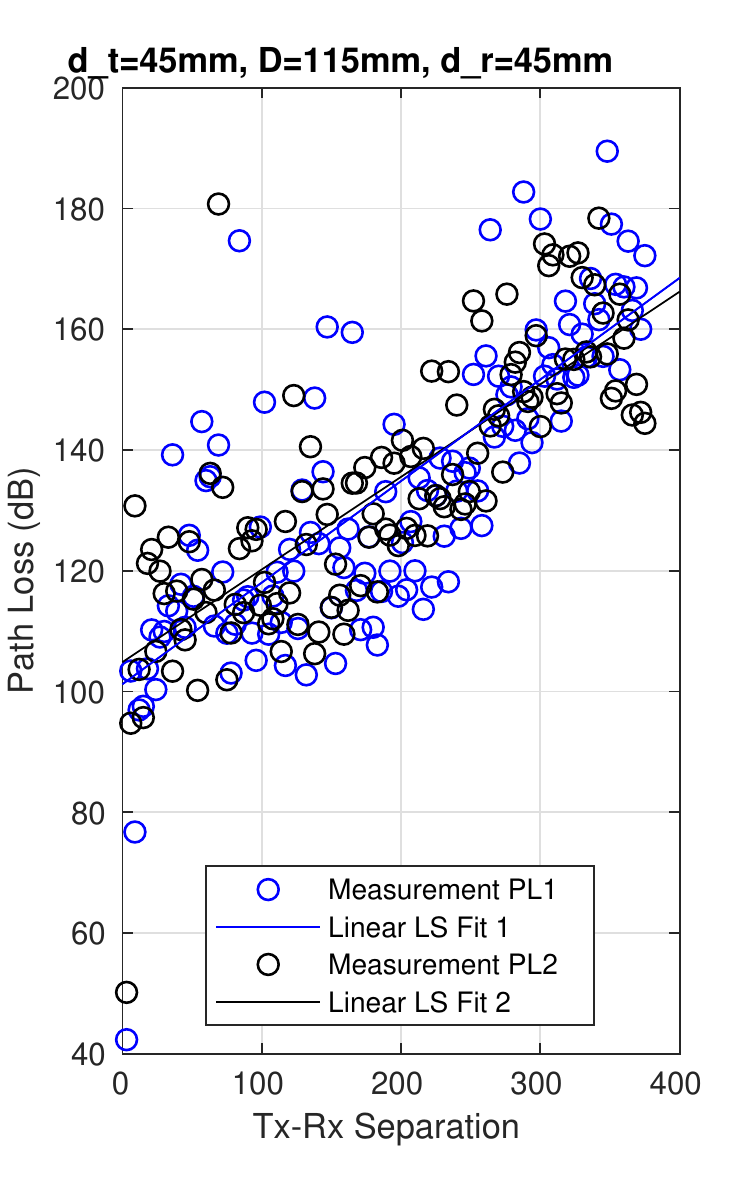}}
\subfloat[\label{fig:pdp}]{\includegraphics[width=0.49\linewidth]{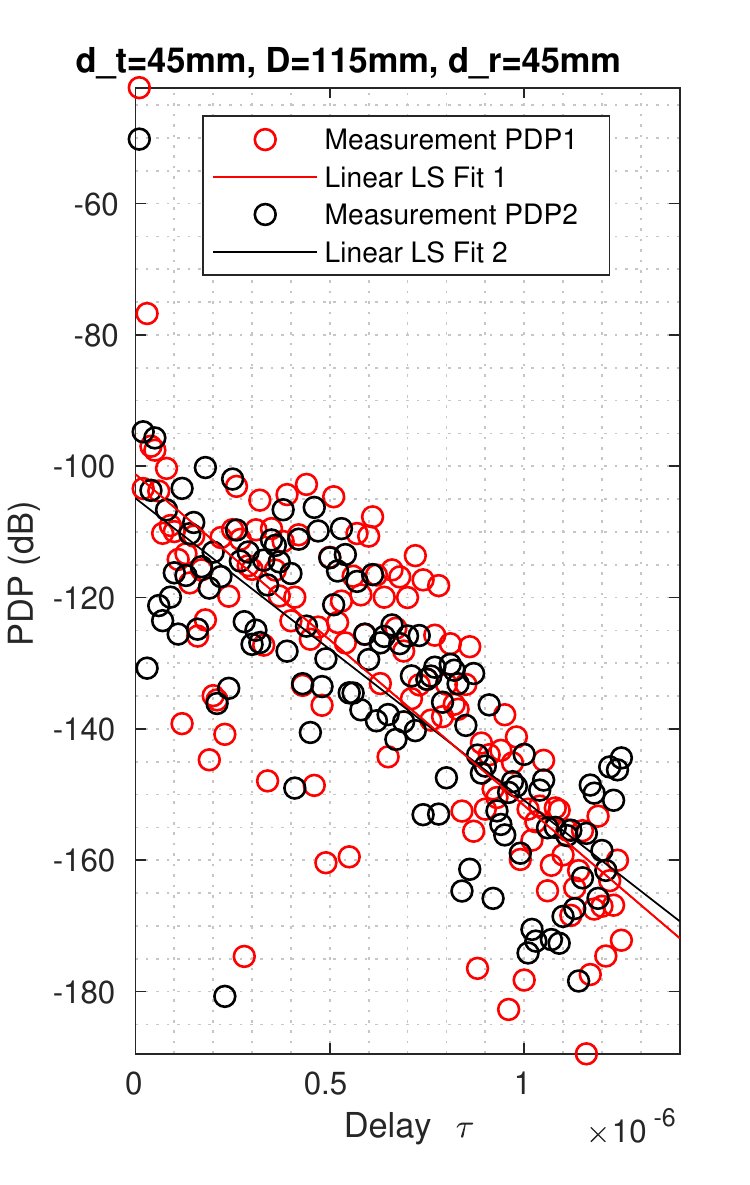}}
    \caption{Linear fitting of the channel parameters (a) path loss and (b) power delay profile.}
    \label{fig:pl_pdp}
\end{figure}

The path loss is important for the design of wireless links and is calculated by inverting the path gains as discussed in \cite{kim2013large}. The path loss can be expressed as:
\begin{equation}
    P_L(d)_{dB}=PL(d_0)_{dB}+10\alpha\log_{10}\left(\frac{d}{d_0} \right)+X_{\sigma},
     \label{eq:pl}
\end{equation}
where $\alpha$ represents path loss exponent. The $X_{\sigma}$ represents the attenuation caused by shadowing, and it follows a Gaussian distribution with zero-mean and standard deviation $\sigma$. We indicate $P_L(d_0)_{dB}$ as the path loss at reference distance $d_0$:
\begin{equation}
    PL(d_0)=20\log_{10}\left(\frac{4\pi d_0}{\lambda} \right).
\end{equation}
The path-loss model is obtained by performing the least-square  fitting of the measured data. Linear fitting gives the intercept and slope of the measured data. In this way, similarly to (\ref{eq:pl}), a statistical path loss model with different path loss exponents and shadowing variable is obtained. 
Figure \ref{fig:pl_pdp} shows the path loss of two received USRP channels.

\subsection{Delay Spread}
The delay spread is a useful parameter that gives the amount of dispersion a wireless channel undergoes in a physical propagation environments. It is measured by selecting a threshold in dB above the noise floor or below the maximum peak of the power delay profile (PDP).  Therefore, the number of MPCs received is strictly function of the threshold and dynamic range of the receiver. The delay spread is given by the following expression \cite{rappaport1996wireless}:
\begin{equation}
    \tau_{RMS}=\sqrt{\frac{\sum_k P(\tau_k)\tau_k^2}{\sum_k P(\tau_k)}-\frac{\sum_k P(\tau_k)\tau_k}{\sum_k P(\tau_k)}}, 
\end{equation}
or more compactly by:
\begin{equation}
    \tau_{RMS}=\sqrt{\overline{\tau^2}-\overline{\tau}},
\end{equation}
where $P(\tau_{k})$ is $k$th power of the multi-path component (MPC) arriving at $\tau_k$, $\overline{\tau}$ is mean excess delay, and $\overline{\tau^2}$ is second central moment of the PDP, with:
\begin{equation}
    \overline{\tau}=\frac{\sum_k P(\tau_k)\tau_k}{\sum_k P(\tau_k)}.
\end{equation}
Figure \ref{fig:pdp} shows the PDP of the two orthogonal MIMO channels.

\subsection{Rician Factor}
The Rician factor is the ratio between the strong line of sight (LOS) component and the scattered NLOS components. Various methods are available in the literature for calculation of Rician factor from narrowband and wideband measurements. Rician factor is helpful to know in presence of rich multipath scattering in the radio propagation environment, especially if there is no direct coupling between transmit and received antennas. Specifically, this is useful when the measurements are performed in a reverberating environment such as an electromagnetic RC or metal enclosures. The Rician factor is ratio of direct power $P_d$ and scattered power $P_s$ component \cite{holloway2006use}\cite{sanchez2010emulation}\cite{lemoine2009advanced,Mariani2020}:
\begin{equation}
    K=\frac{P_d}{P_s}=\frac{|\overline{S21}|^2}{\overline{|S21-\overline{S21}|^2}}.
\end{equation}
   
\begin{figure}
    \centering
    \includegraphics[width=\linewidth]{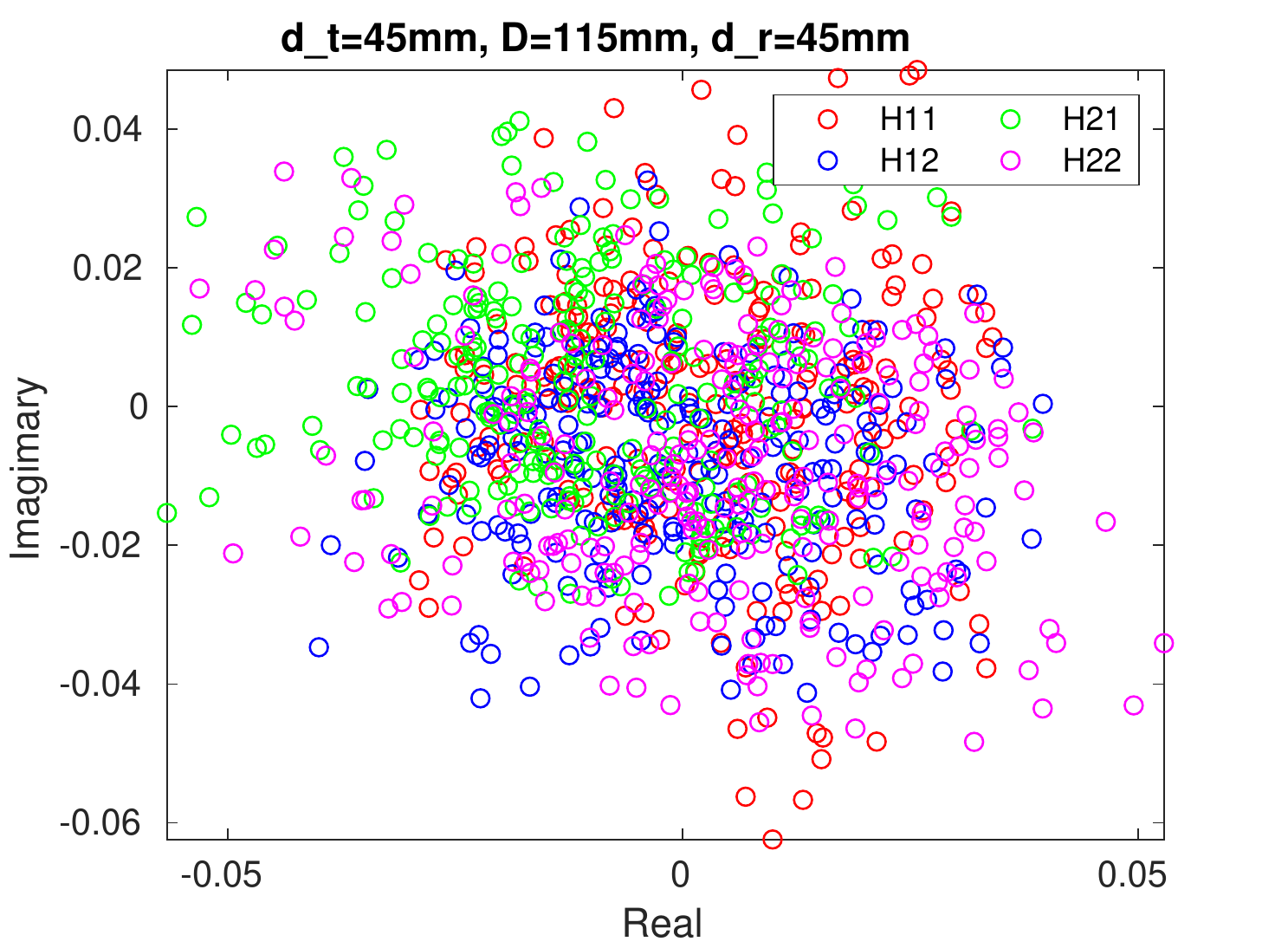}
    \caption{Complex cloud of MIMO sub-channels centered around zero.}
    \label{fig:complex_cloud}
\end{figure}
Figure \ref{fig:complex_cloud} shows the scatter plot of sub-channel elements of the MIMO channel matrix. It is shown that the real and imaginary values are grouped around the zero-axis. This indicates that sufficient scattering occurs and a negligible direct coupling or LOS component.

\subsection{Distribution of the Channel Capacity}
\begin{figure}
        \centering
        \includegraphics[width=\linewidth]{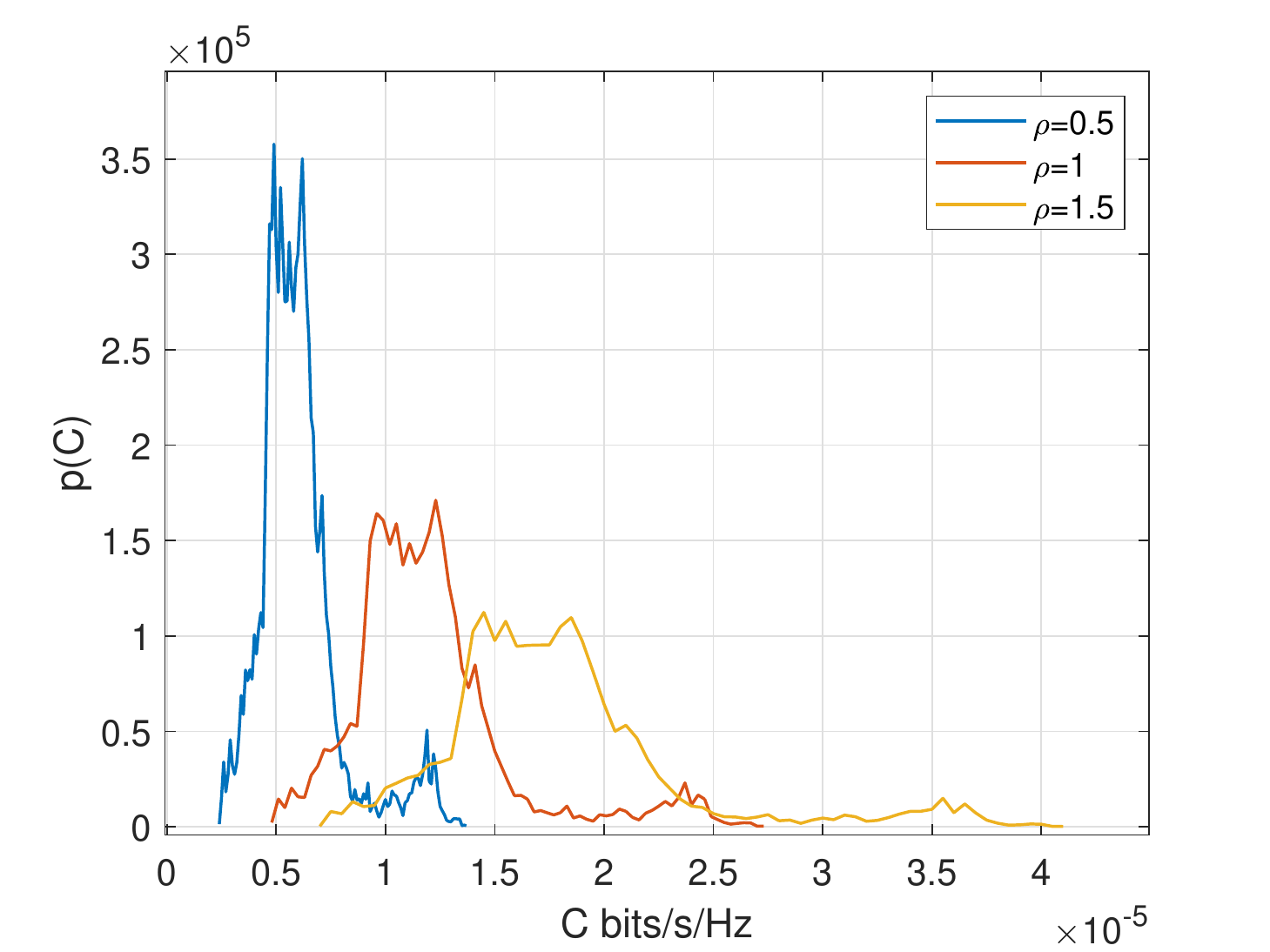}
        \caption{Distribution for the channel capacity with no cones in $25$mm $\times$ $25$ mm metal enclosure.}
        \label{fig:pdf_channel1}
        \end{figure}

\begin{figure}
          \centering
        \includegraphics[width=\linewidth]{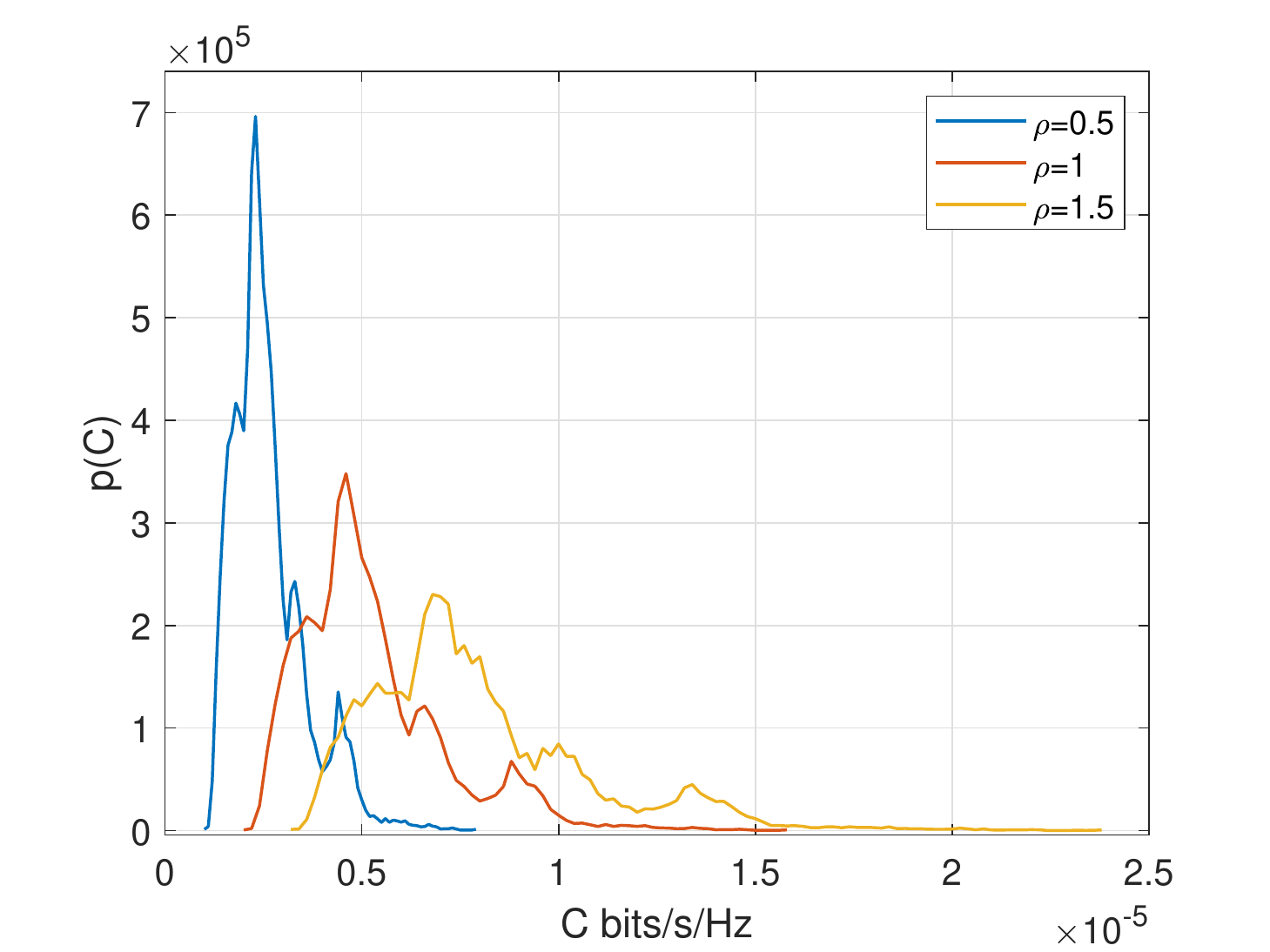}
        \caption{Distribution for the channel capacity with $2$ cones in $25$ mm$\times$ $25$mm metal enclosure.}
        \label{fig:pdf_channel2}
        \end{figure}

\begin{figure}
          \centering
        \includegraphics[width=\linewidth]{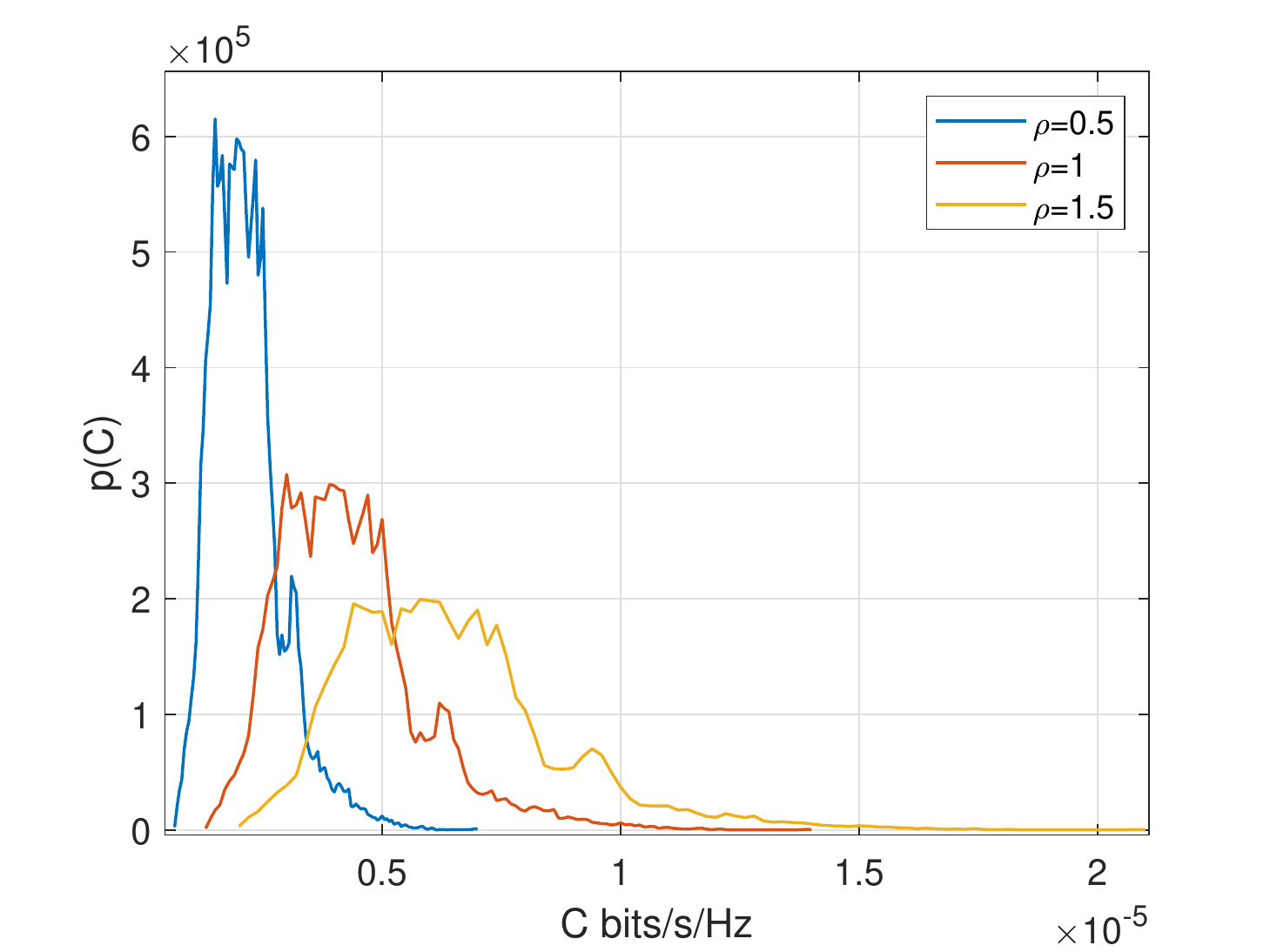}
        \caption{Distribution for the channel capacity with $4$ cones in $25$mm$\times$ $25$mm metal enclosure.}
        \label{fig:pdf_channel3}
        \end{figure}

\begin{figure}
          \centering
        \includegraphics[width=\linewidth]{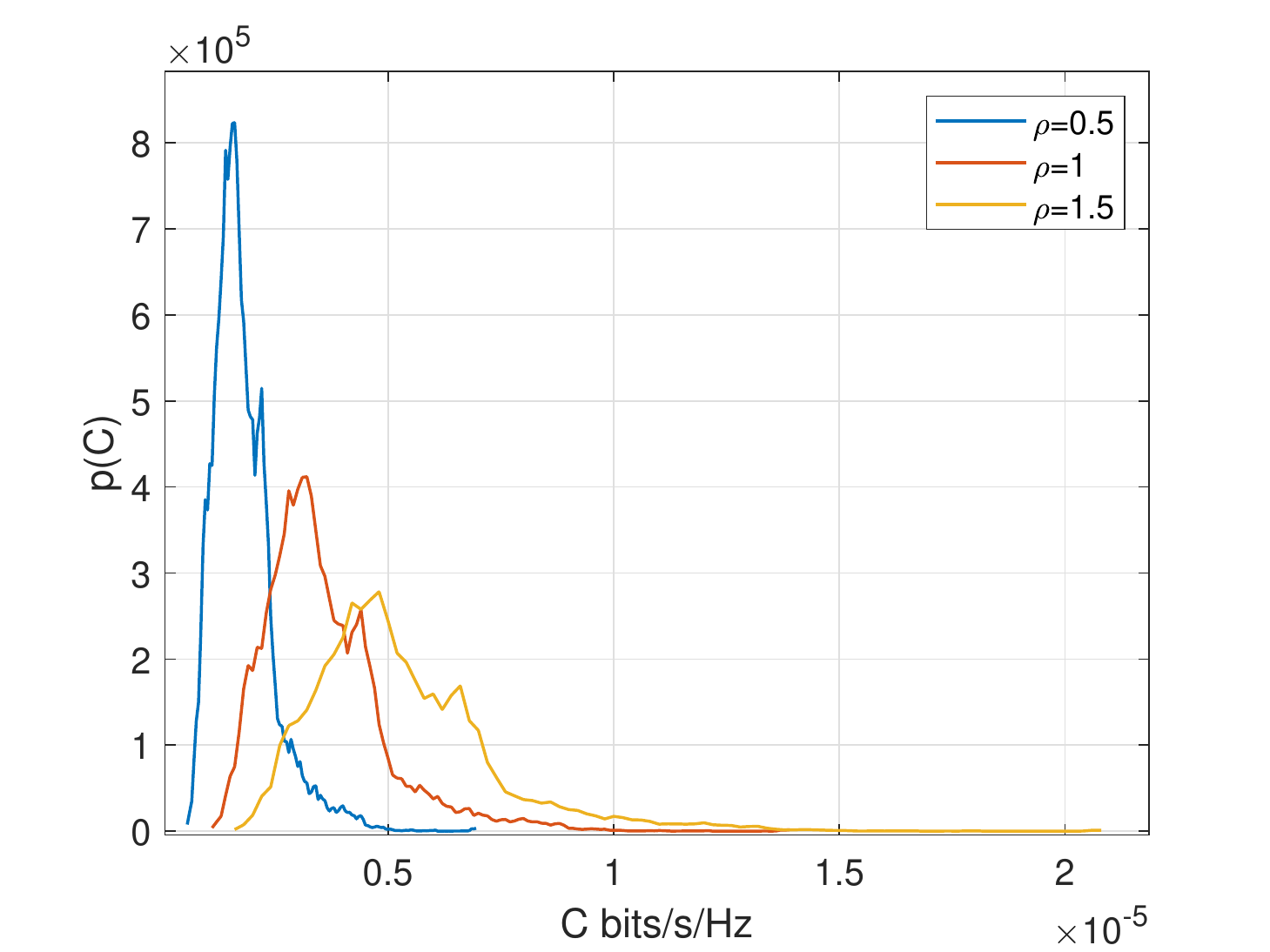}
        \caption{Distribution for the channel capacity with $8$ cones in $25$mm$\times$ $25$mm metal enclosure.}
        \label{fig:pdf_channel4}
    \end{figure}

\begin{figure}
        \centering
        \includegraphics[width=\linewidth]{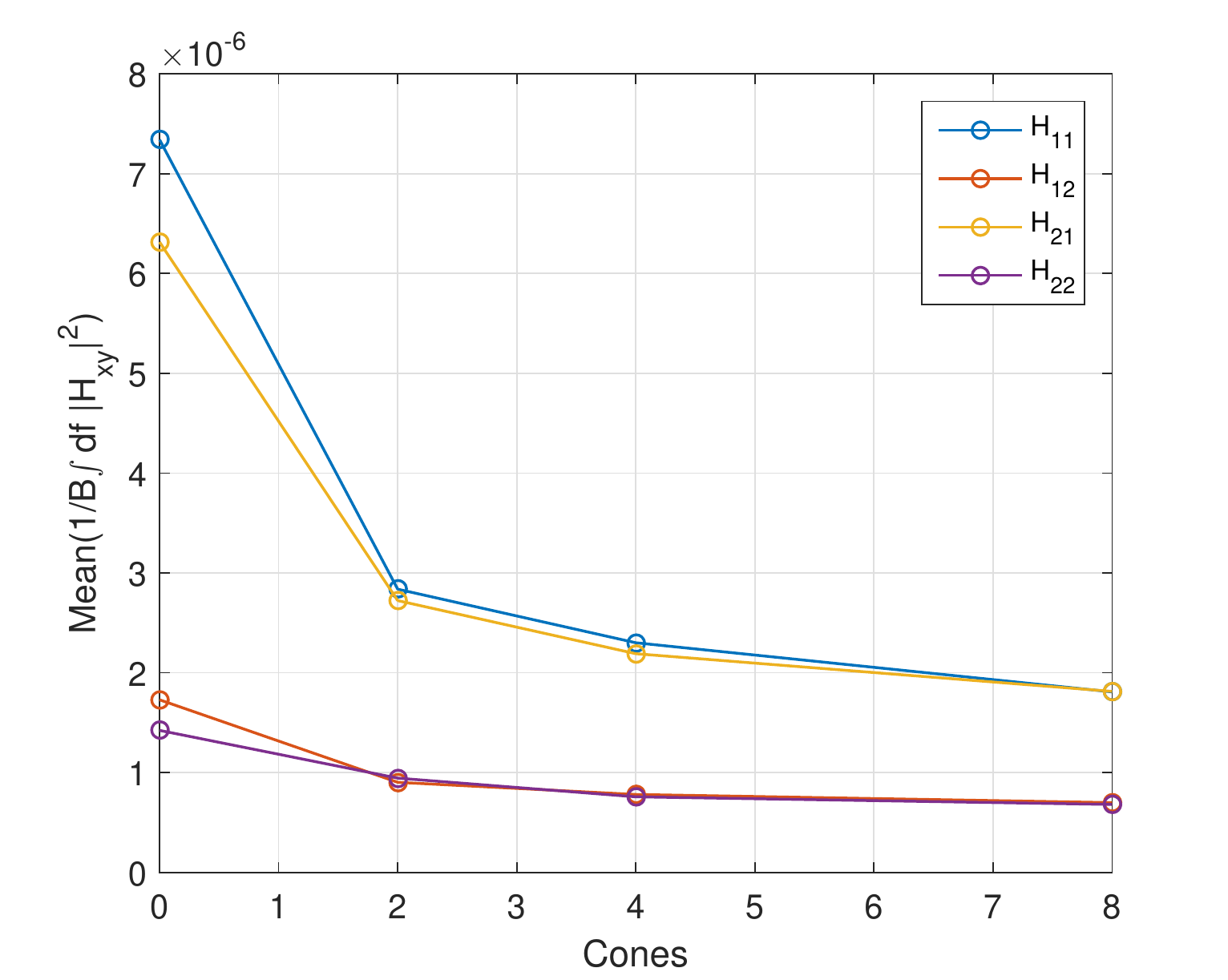}
        \caption{Mean of Eq(\ref{Hxy2stat})  for different cones settings in the $25$mm$\times$ $25$mm.}
        \label{fig:mean_channel1}
        \end{figure}

\begin{figure}
        \centering
        \includegraphics[width=\linewidth]{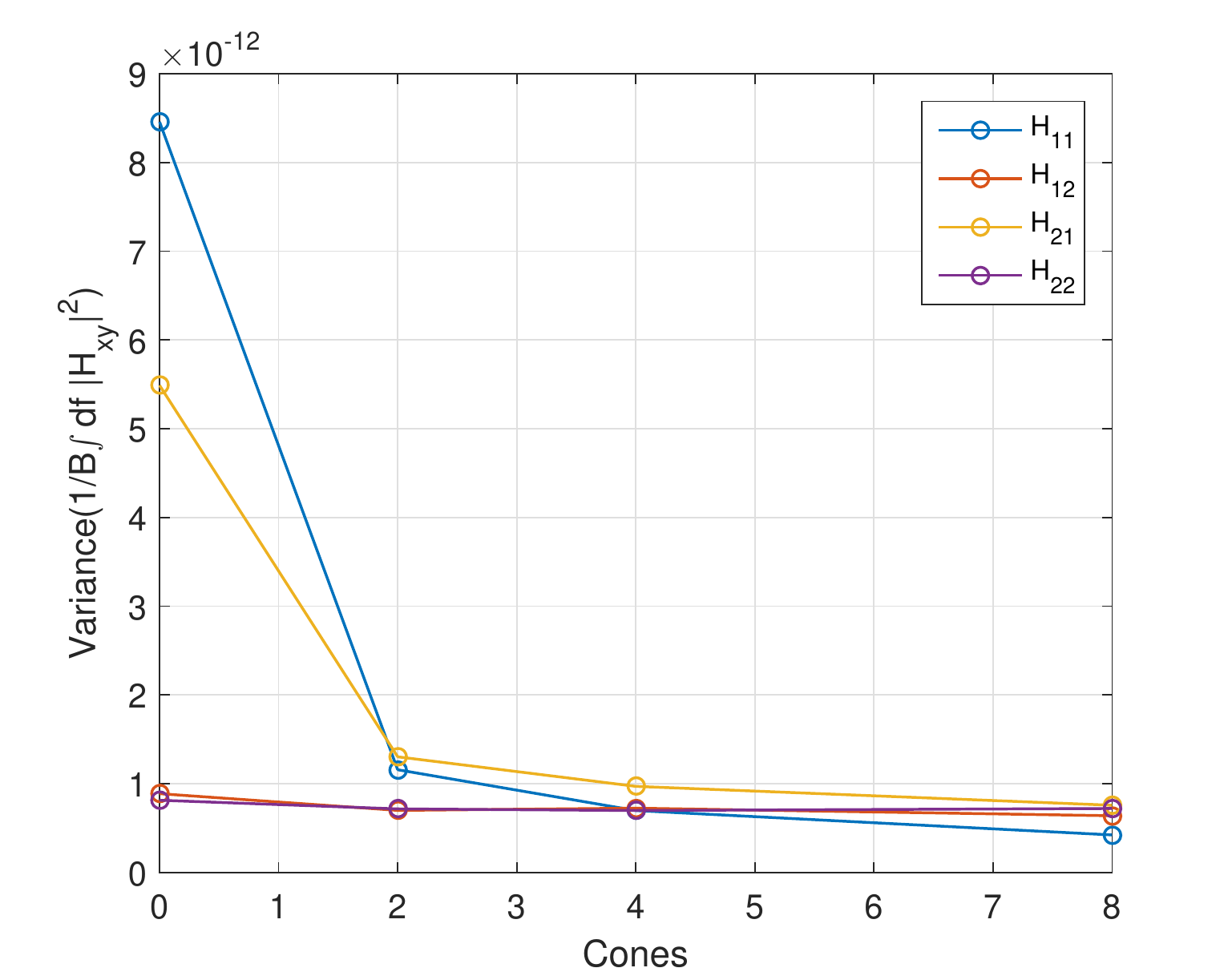}
        \caption{Variance of Eq(\ref{Hxy2stat})  for different cones settings in the $25$mm$\times$ $25$mm.}
        \label{fig:var_channel1}
        \end{figure}

We now turn our attention to the statistics of the channel capacity $C$ starting from measured data. 
Figures \ref{fig:pdf_channel1},\ref{fig:pdf_channel2},\ref{fig:pdf_channel3}, and \ref{fig:pdf_channel4}) show the probability density function (PDF) of the channel capacity for chamber in empty and selected loaded conditions. The empirical distributions show isolated and dominant peaks and, therefore, are multimodal. Moreover, intervals separating consecutive peaks is controlled by the amount on losses while multimodality seems to disappear as the losses increase. This behaviour, far from a Gaussian distribution, is hardly analytically accessible. 
Furthermore, idealised assumptions commonly used for statistical channel modelling, as well as fading characterization, need to be revisited from a theoretical point of view.  However, for sufficiently small $\rho$, the range of $C$ allows one to simplify the investigations, i.e. typical realization of $C$ are given by: 
\begin{equation}\label{eq:Capprox1}
C\approx\frac{\rho }{2 B\ln 2}\int_{f_{min}}^{f_{max}} df  H(f) H(f)^\dagger.
\end{equation}
Therefore, one is left with the statistics of $ H(f)H(f)^\dagger=\sum_{ij=1}^{2}|H_{ij}|^2(f)$. In Figs. \ref{fig:mean_channel1},\ref{fig:var_channel1}, we evaluate the first statistics of:
\begin{equation}\label{Hxy2stat}
    \frac{1}{B}\int_{f_{min}}^{f_{max}}df |H_{xy}|^2(f).
\end{equation}
Experiments confirm the monotonicity of its mean and variance as function of absorption \cite{Gradoni2020}. The appearance of two curves in Figs.\ref{fig:mean_channel1},\ref{fig:var_channel1} is due to the selective excitement of the transmit antennas while measuring the channel coefficient at two receiver antennas. 


\section{Conclusion}
A thorough statistical characterization of an SDR-based 2x2 MIMO wireless system operating in a mode-stirred metal enclosure has been performed and discussed. The MIMO channel measurements extracted from the USRP-based tesdbed show the formation of two parallel MIMO channels. In particular, the creation of an appropriate MIMO channel transfer matrix is supported by the analysis of key performance indicators under severe dynamical multi-path conditions emulated by a mechanical mode-stirrer process. Starting from measured complex-valued transfer function of the MIMO wireless channel, PDP, path loss, coherence bandwidth, and Rician factor are obtained following definitions grounded on statistics RC theory. The two channels show substantial gain but are unbalanced, suggesting that equally strong channels can be achieved if decoupling and matching networks are used at both transmit and receive sides, even in presence of moderate losses. Additionally, MIMO channel capacity results show strong non-Gaussianity which is function of losses introduced in the enclosure. In particular, multi-modal probability distributions are observed and multi-modality is mitigated by increasing the losses of the enclosure. Both the ergodic channel capacity and the associated standard deviation scale monotonically with losses.

\bibliographystyle{IEEEtran}
\bibliography{wc2c.bib} 

\begin{thebibliography}{10}
\providecommand{\url}[1]{#1}
\csname url@samestyle\endcsname
\providecommand{\newblock}{\relax}
\providecommand{\bibinfo}[2]{#2}
\providecommand{\BIBentrySTDinterwordspacing}{\spaceskip=0pt\relax}
\providecommand{\BIBentryALTinterwordstretchfactor}{4}
\providecommand{\BIBentryALTinterwordspacing}{\spaceskip=\fontdimen2\font plus
\BIBentryALTinterwordstretchfactor\fontdimen3\font minus
  \fontdimen4\font\relax}
\providecommand{\BIBforeignlanguage}[2]{{%
\expandafter\ifx\csname l@#1\endcsname\relax
\typeout{** WARNING: IEEEtran.bst: No hyphenation pattern has been}%
\typeout{** loaded for the language `#1'. Using the pattern for}%
\typeout{** the default language instead.}%
\else
\language=\csname l@#1\endcsname
\fi
#2}}
\providecommand{\BIBdecl}{\relax}
\BIBdecl

\bibitem{chen2007inter}
Z.~M. Chen and Y.~Zhang, ``Inter-chip wireless communication channel:
  Measurement, characterization, and modeling,'' \emph{IEEE Transactions on
  Antennas and Propagation}, vol.~55, pp. 978--986, 2007.

\bibitem{zajic2018modeling}
A.~Zajic and P.~Juyal, ``Modeling of thz chip-to-chip wireless channels in
  metal enclosures,'' in \emph{12th European Conference on Antennas and
  Propagation (EuCAP 2018)}.\hskip 1em plus 0.5em minus 0.4em\relax IET, 2018,
  pp. 1--5.

\bibitem{fu2019300}
J.~Fu, P.~Juyal, and A.~Zaji{\'c}, ``300 ghz channel characterization of
  chip-to-chip communication in metal enclosure,'' in \emph{2019 13th European
  Conference on Antennas and Propagation (EuCAP)}.\hskip 1em plus 0.5em minus
  0.4em\relax IEEE, 2019, pp. 1--5.

\bibitem{kim2016300}
S.~Kim and A.~Zaji{\'c}, ``300 ghz path loss measurements on a computer
  motherboard,'' in \emph{2016 10th European Conference on Antennas and
  Propagation (EuCAP)}.\hskip 1em plus 0.5em minus 0.4em\relax IEEE, 2016, pp.
  1--5.

\bibitem{karedal2007characterization}
J.~Karedal, A.~P. Singh, F.~Tufvesson, and A.~F. Molisch, ``Characterization of
  a computer board-to-board ultra-wideband channel,'' \emph{IEEE Communications
  letters}, vol.~11, no.~6, pp. 468--470, 2007.

\bibitem{redfield2011understanding}
S.~Redfield, S.~Woracheewan, H.~Liu, P.~Chiang, J.~Nejedlo, and R.~Khanna,
  ``Understanding the ultrawideband channel characteristics within a computer
  chassis,'' \emph{IEEE Antennas and Wireless Propagation Letters}, vol.~10,
  pp. 191--194, 2011.

\bibitem{khademi2015channel}
S.~Khademi, S.~P. Chepuri, Z.~Irahhauten, G.~J. Janssen, and A.-J. van~der
  Veen, ``Channel measurements and modeling for a 60 ghz wireless link within a
  metal cabinet,'' \emph{IEEE Transactions on Wireless Communications},
  vol.~14, no.~9, pp. 5098--5110, 2015.

\bibitem{ohira2011experimental}
M.~Ohira, T.~Umaba, S.~Kitazawa, H.~Ban, and M.~Ueba, ``Experimental
  characterization of microwave radio propagation in ict equipment for wireless
  harness communications,'' \emph{IEEE Transactions on Antennas and
  Propagation}, vol.~59, no.~12, pp. 4757--4765, 2011.

\bibitem{nakamoto2013wireless}
N.~Nakamoto, H.~Ban, T.~Oka, S.~Kitazawa, K.~Kobayashi, N.~Kikuchi,
  H.~Hatamoto, S.~Shimizu, and M.~Hara, ``Wireless harness inside ict
  equipments,'' in \emph{2013 15th International Conference on Advanced
  Communications Technology (ICACT)}.\hskip 1em plus 0.5em minus 0.4em\relax
  IEEE, 2013, pp. 135--143.

\bibitem{Micheli2021}
``Mimo 4x4 vs. mimo 2x2 performance assessment of a real life lte base station
  in a reverberation chamber,'' \emph{AEU - International Journal of
  Electronics and Communications}, vol. 129, p. 153500, 2021.

\bibitem{rayess2017antennas}
W.~Rayess, D.~W. Matolak, S.~Kaya, and A.~K. Kodi, ``Antennas and channel
  characteristics for wireless networks on chips,'' \emph{Wireless Personal
  Communications}, vol.~95, no.~4, pp. 5039--5056, 2017.

\bibitem{chuah2002capacity}
C.-N. Chuah, D.~N.~C. Tse, J.~M. Kahn, and R.~A. Valenzuela, ``Capacity scaling
  in mimo wireless systems under correlated fading,'' \emph{IEEE Transactions
  on Information theory}, vol.~48, no.~3, pp. 637--650, 2002.

\bibitem{molisch2002capacity}
A.~F. Molisch, M.~Steinbauer, M.~Toeltsch, E.~Bonek, and R.~S. Thoma,
  ``Capacity of mimo systems based on measured wireless channels,'' \emph{IEEE
  Journal on selected areas in communications}, vol.~20, no.~3, pp. 561--569,
  2002.

\bibitem{tse2005fundamentals}
D.~Tse and P.~Viswanath, \emph{Fundamentals of wireless communication}.\hskip
  1em plus 0.5em minus 0.4em\relax Cambridge university press, 2005.

\bibitem{Phang2018}
S.~{Phang}, M.~T. {Ivrlač}, G.~{Gradoni}, S.~C. {Creagh}, G.~{Tanner}, and
  J.~A. {Nossek}, ``Near-field mimo communication links,'' \emph{IEEE
  Transactions on Circuits and Systems I: Regular Papers}, vol.~65, no.~9, pp.
  3027--3036, 2018.

\bibitem{TannerNEMF}
G.~T. et~al., \emph{FETOpen Project NEMF21}, 2018,
  https://cordis.europa.eu/project/id/664828/results.

\bibitem{chen2011estimation}
X.~Chen, P.-S. Kildal, and S.-H. Lai, ``Estimation of average rician k-factor
  and average mode bandwidth in loaded reverberation chamber,'' \emph{IEEE
  Antennas and Wireless Propagation Letters}, vol.~10, pp. 1437--1440, 2011.

\bibitem{delangre2008delay}
O.~Delangre, P.~De~Doncker, M.~Lienard, and P.~Degauque, ``Delay spread and
  coherence bandwidth in reverberation chamber,'' \emph{Electronics letters},
  vol.~44, no.~5, pp. 328--329, 2008.

\bibitem{kim2013large}
M.~Kim, Y.~Konishi, Y.~Chang, and J.-i. Takada, ``Large scale parameters and
  double-directional characterization of indoor wideband radio multipath
  channels at 11 ghz,'' \emph{IEEE Transactions on Antennas and Propagation},
  vol.~62, no.~1, pp. 430--441, 2013.

\bibitem{rappaport1996wireless}
T.~S. Rappaport \emph{et~al.}, \emph{Wireless communications: principles and
  practice}.\hskip 1em plus 0.5em minus 0.4em\relax prentice hall PTR New
  Jersey, 1996, vol.~2.

\bibitem{holloway2006use}
C.~L. Holloway, D.~A. Hill, J.~M. Ladbury, P.~F. Wilson, G.~Koepke, and
  J.~Coder, ``On the use of reverberation chambers to simulate a rician radio
  environment for the testing of wireless devices,'' \emph{IEEE transactions on
  antennas and propagation}, vol.~54, no.~11, pp. 3167--3177, 2006.

\bibitem{sanchez2010emulation}
J.~D. S{\'a}nchez-Heredia, J.~F. Valenzuela-Vald{\'e}s, A.~M.
  Mart{\'\i}nez-Gonz{\'a}lez, and D.~A. Sanchez-Hernandez, ``Emulation of mimo
  rician-fading environments with mode-stirred reverberation chambers,''
  \emph{IEEE Transactions on Antennas and Propagation}, vol.~59, no.~2, pp.
  654--660, 2010.

\bibitem{lemoine2009advanced}
C.~Lemoine, P.~Besnier, and M.~Drissi, ``Advanced method for estimating
  direct-toscattered ratio of rician channel in reverberation chamber,''
  \emph{Electronics Letters}, vol.~45, no.~4, pp. 194--196, 2009.

\bibitem{Mariani2020}
V.~M. {Primiani}, M.~{Barazzetta}, L.~{Bastianelli}, D.~{Micheli}, E.~{Moglie},
  R.~{Diamanti}, and G.~{Gradoni}, ``Reverberation chambers for testing
  wireless devices and systems,'' \emph{IEEE Electromagnetic Compatibility
  Magazine}, vol.~9, no.~2, pp. 45--55, 2020.

\bibitem{Gradoni2020}
G.~{Gradoni}, M.~{Richter}, S.~{Phang}, S.~B. {Fedeli}, U.~{Kuhl},
  O.~{Legrand}, and A.~{Ishimaru}, ``Statistical model for mimo propagation
  channel in cavities and random media,'' in \emph{2020 XXXIIIrd General
  Assembly and Scientific Symposium of the International Union of Radio
  Science}, 2020, pp. 1--4.

\end{thebibliography}
\end{document}